\documentclass[journal]{IEEEtran}
\usepackage{cite}

\ifCLASSINFOpdf
   \usepackage[pdftex]{graphicx}
\else
\fi
\usepackage{amsmath}
\usepackage{amsfonts}
\usepackage[safe]{tipa}
\usepackage{xurl}
\begin{document}
%
\title{Interpreting intermediate convolutional layers of generative CNNs trained on waveforms}
%
%
%

\author{Ga\v{s}per Begu\v{s} and Alan Zhou
\thanks{This work was supported in part by a grant to new faculty at UC Berkeley.}
\thanks{Ga\v{s}per Begu\v{s} is with the Department of Linguistics, University of California, Berkeley, 1203 Dwinelle Hall \#2650,
Berkeley, CA 94720-2650, United States (e-mail: begus@berkeley.edu). \newline Alan Zhou is with the Department of Cognitive Science, Johns Hopkins University, 237 Krieger Hall, 3400 N.~Charles Street, Baltimore, MD 21218, United States  (e-mail: azhou23@jhu.edu).}
\thanks{This paper has supplementary downloadable material available at http://ieeexplore.ieee.org., provided by the author. The material includes Supplementary Materials (pdf). Contact begus@berkeley.edu for further questions about this work.}}

%
%

\markboth{}%
{Interpreting intermediate convolutional layers of generative CNNs trained on waveforms}
%



\maketitle



\begin{abstract}
This paper presents a technique to interpret and visualize intermediate layers in generative CNNs trained on raw speech data in an unsupervised manner. We argue that averaging over feature maps after ReLU activation in each transpose convolutional layer yields interpretable time-series data. This technique allows for acoustic analysis of intermediate layers that parallels the acoustic analysis of human speech data: we can extract F0, intensity, duration, formants, and other acoustic properties from intermediate layers in order to test where and how CNNs encode various types of information. We further combine this technique with linear interpolation of a model's latent space to show a causal relationship between individual variables in the latent space and activations in a model's intermediate convolutional layers. In particular, observing the causal effect between linear interpolation and the resulting changes in intermediate layers can reveal how individual latent variables get transformed into spikes in activation in intermediate layers. We train and probe internal representations of two models --- a bare WaveGAN architecture and a ciwGAN extension which forces the Generator to output informative data and results in the emergence of linguistically meaningful representations. Interpretation and visualization is performed for three basic acoustic properties of speech: periodic vibration (corresponding to vowels), aperiodic noise vibration (corresponding to fricatives), and silence (corresponding to stops). The proposal also allows testing of higher-level morphophonological alternations such as reduplication (copying). In short, using the proposed technique, we can analyze how linguistically meaningful units in speech get encoded in each convolutional layer of a generative neural network.
\end{abstract}


%

\IEEEpeerreviewmaketitle

\section{Introduction}
\label{intro}

How deep convolutional neural networks learn their internal representations is one of the central questions in machine learning. The vast majority of work on this topic is centered on the visual domain. Here, we propose a technique to visualize and interpret the intermediate layers of deep convolutional neural networks trained on speech in an unsupervised manner. There are several advantages of interpreting intermediate layers in convolutional neural networks (CNNs) that are trained on speech over those trained on visual data. 

First, humans process speech by discretizing the continuous physical properties of sound into discrete mental representations called \emph{phonemes}. A long tradition of scientific study of phonetics and phonology \cite{hulst13,macmahon13} has resulted in a relatively good understanding of how humans represent continuous properties in speech with such discrete units. A process reminiscent of human phonology emerges in unsupervised CNNs: they learn to represent the continuous space of spoken language with discretized representations \cite{begus19,begusCiw}. 

Second, speech data contains multiple local and non-local dependencies with different degrees of computational complexity that are well-documented and well-understood. For example, changing or adding a single sound to a word can result in a change in meaning. The English word \emph{pit} [\textipa{"p\super hIt}] has a different meaning from \emph{spit} [\textipa{"spIt}]. Two processes occur here. First, the addition of the sound [s] changes the meaning of the word. Second, the stop consonant is produced with aspiration (puff of air marked by \textipa{\super h}) in the first word with no preceding [s], but without aspiration in the second word with preceding [s]. This contextually conditioned complementary distribution (between [\textipa{p\super h}] and [\textipa{p}]) is computationally simple, but this is not true for all processes in human speech. For example, many natural languages feature an identity-based process called reduplication which requires phonological material to be copied from the output. A reduplicated form of the base [\textipa{para}] is  [\textipa{papara}], where the first consonant and the first vowel [pa] in the base [para] are repeated (copied), which results in [\textipa{papara}]. Reduplication is on a higher complexity level than other phonological processes on the Chomsky hierarchy; it is more than context-free when most other phonological processes are regular\cite{gasser93,marcus99,brugiapaglia20,savitch89,dolatian20,wang21}. Specifically, it is a non-concatenative process that requires learners to copy phonological material from the base: [pa] is the prefix only for the bases starting with [pa] such as [para]. For other bases (such as [\textipa{tara}] or [\textipa{mura}]), the reduplication morphemes that fulfill exactly the same function are substantially different: [\textipa{ta}] and [\textipa{mu}], which makes learning more challenging than simple concatenative patterns. We can use these well-understood dependencies with different degrees of computational complexity to test what internal representations are learned from raw continuous data by CNNs and how they are learned. We are also able to test which acoustic properties get encoded at each convolutional layer.

Finally, an advantage of interpreting CNNs trained on speech is that behavioral and acquisitional data is easy to obtain for speech. We can directly compare developmental stages in child language acquisition with stages of CNNs trained on speech \cite{begus19}, or use the same data to train human subjects and CNNs \cite{begusLocal}. Outputs of the proposed technique can also be directly compared with human neuroimaging data, which contains time-series data of electrical activity in different parts of the brain recorded with various neuroimaging techniques.

\section{Prior work}

Visualizing convolutional layers is performed primarily on models trained on visual data \cite{zeiler14}, with considerably less work focused on the visualization of convolutional layers of the models trained on speech \cite{huang15,krug19,muckenhirn19,chowdhury20}.  For example, work on unsupervised models such as generative adversarial networks (GANs) has primarily been carried out on image data, and has been successful in identifying relationships in the latent space \cite{shen2020interpreting}, as well as intermediate representations of various generated classes  \cite{bau2018gan}. These approaches often leverage techniques specific to the visual domain, such as attribute prediction and image segmentation.

\subsection{Interpreting models trained on speech}

Substantially less work exists on interpreting convolutional layers trained on speech, and the majority of this work operates on supervised models. Many proposals focus on interpreting and visualizing filters. The SincNet proposal \cite{ravanelli19} visualizes filters, and by imposing restrictions on filters, achieves better performance on an ASR task compared to unrestricted CNNs. Huang et al.~\cite{huang15} likewise focus on visualizing filters of convolutional layers from supervised models trained for ASR tasks. \cite{krug18introspection,krug18,krug19} make use of activation maps of convolutions on spectrogram inputs, using them to compute neuron activation profiles. The proposed techniques can highlight important regions for ASR tasks in CNNs, but focus more on individual neuron activations than intermediate representations. Millet and King \cite{millet21} analyze activations in deep neural networks and correlate them with fMRI data.

Palaz et al.~\cite{palaz13,palaz15,palaz19}, Muckenhirn et al.~\cite{muckenhirn17,muckenhirn18}, and Golik et al.~\cite{golik15} also analyze learned filters at different convolutional layers. Muckenhirn et al.~\cite{muckenhirn18} analyze filters of CNN models for ASR tasks, but trained on raw waveforms. They also visualize estimated F0 contours based on filters in the first convolutional layer \cite{muckenhirn18}. Analysis of the filters can, for example, reveal which frequency bands various filters target.  This can in turn reveal what types of acoustic data are encoded at which convolutional layers. However, the proposed techniques yield less directly interpretable outputs.  For example, this technique does not allow a  direct analysis of waveforms from individual convolutional layer that directly correspond to some phonetic element in the final output layer.

Muckenhirn et al.~\cite{muckenhirn19} propose a gradient-based visualization technique for CNNs trained on raw waveforms (based on \cite{springenberg15}) which yields relevance maps from the input signal that can be acoustically analyzed (a similar proposal that uses relevance maps is in \cite{chowdhury20}). Their models are trained on supervised tasks: phone or speaker identification. Similar to our technique, their proposal enables analysis of acoustic properties (such as formant values and F0) in CNNs on a time-series data. Their method, however, does not focus on analyzing which acoustic representations are encoded at each layer, focusing instead on the most relevant parts of the input signal. Their supervised model also lacks the ability to test effects of individual latent variables on convolutional layers. Additionally, they focus on spectral analyses as they argue that  ``[v]isualization in the time domain does not bring much insights into what important characteristics are extracted by the network because the results are difficult to interpret, as we do not have any visual cues as in the case of images'' \cite[p.~2346]{muckenhirn19}. This paper argues that averaged ReLU activations of feature maps combined with manipulation and linear interpolation of individual linguistically meaningful latent variables yield highly interpretable time-series data. Koumura et al. \cite{Koumura2019Jul} also examines a CNN trained on raw waveforms, taking inspiration from single neuron recordings to examine the activations of individual units. They examine the synchrony of individual activations to an input stimulus, and take averages across time rather than across layers. Harwath and Glass \cite{harwath19} take the L2 norm of activation maps in spectrograms and perform a PCA analysis of their outputs. Their work focuses on phoneme transition marking in one convolutional layer. To our knowledge, none of the proposals test the causal effect between latent variables and intermediate convolutional layers or probe representations in a generative model (which brings several advantages outlined below).

\subsection{Our approach} 
 
Here, we propose a different approach for interpreting intermediate convolutional layers from the existing proposals outlined above. By interpretability, we mean the ability to analyze how meaningful units in data are represented in intermediate convolutional layers. We propose a set of techniques that enables testing predictions such as what acoustic properties are encoded at what layer (and how) in a decoder (Generator) network. Rather than analyzing convolutional layers in a supervised model or analyzing filters, our proposal focuses on the activations of intermediate transpose convolutions of a Generator network that was trained on speech in a GAN framework. Whereas traditional convolutions are usually used to downsample preexisting data into lower-dimensional representations, transpose convolutions work in reverse, upsampling from a low-dimensional latent representation in order to generate new data. This framework causes some key differences in the structure of our intermediate layers, with the highest-level representations appearing in the deepest layers of the network. However, it also allows for exploration of causal relationships between representations of phonetic units in the latent space and encoding of those units in intermediate convolutional layers.

Our proposal brings several aspects that facilitate the interpretability of the activations of these intermediate convolutional layers. In \cite{begusZhou1}, we propose that averaging over feature maps yields interpretable time series data, but focus exclusively on the classifier network in which the causal relationship between the network's classification output  and intermediate layers cannot be established. \cite{begusZhou1} also focuses only on encoding of words and individual acoustic contrasts (and not phonological processes). Here, we follow the proposal in \cite{begusZhou1}, but focus on the Generator network. We introduce several new approaches to the paradigm:  (i) manipulating and interpolating individual latent variables well beyond training range (based on \cite{begus19}) while visualizing intermediate layers, which enables   (ii) observing the causal relationship between individual variables in the latent space and linguistically meaningful units in intermediate layers; (iii) testing which acoustic properties are encoded at which layer via correlations; and (iv) testing not only encoding of acoustic properties or words, but also of phonological processes and higher-level morphophonological processes such as reduplication. Like in \cite{begusZhou1}, we train the networks in an unsupervised manner and interpret time-series data directly.\footnote{Other proposals also operate with raw waveforms and some also visualize feature maps (see above).} Below we outline why these aspects are important. 

One of the main difficulties with interpreting convolutional layers in supervised ASR models is that it is not trivial to elicit or amplify activations given that the network takes raw data as inputs and outputs some classification, as we can only directly modify the raw data input. We propose an interpretable alternative: we build on a proposal in \cite{begus19} that individual latent variables in a generative model can be manipulated to marginal levels well outside the training range, and that linear interpolation can reveal the causal relationship between individual variables and meaningful linguistic representations in the output and apply these two concepts to the visualization paradigm. The majority of proposals on CNN interpretability, known to the authors, do not manipulate individual latent variables. The generative and unsupervised aspect of the GAN framework (namely WaveGAN and ciwGAN) make this technique possible: we can manipulate the latent space and observe causal effects of individual meaningful variables on intermediate layers.  In other words, we can observe how individual variables with some linguistic function get transformed throughout the convolutional layers while keeping the rest of the latent space $z$ constant.

We interpret and visualize intermediate convolutional layers in a fully unsupervised manner --- in the GAN framework.  The majority of ASR/synthesis models using CNNs are supervised. The advantage of interpreting intermediate layers on unsupervised models is that the final reduced representation layer is not trained on a classification problem with a softmax function, but is connected to uniformly distributed random variables (or a combination of binary and uniformly distributed random variables) that get transformed to data in the output layer. This means that we can analyze self-organization of meaningful representations in intermediate convolutional layers and directly observe effects of individual variables in the latent space on intermediate representations. 

The same technique can also be applied to other zero-resource speech models for unsupervised acoustic word embedding \cite{levin13,dunbar20} (such as autoencoders \cite{chorowski19,kamper19,niekerk20,chung20,chen20,eloff19,shain19,baevski20}), but GANs are chosen because they are unsupervised not only in the encoding task, but also in the generative task and as such even more suitable for generating novel outputs.  Unlike in variational autoencoders (VAEs), the generator of a GAN never directly accesses the training data.  In the GAN architecture, the generation aspect is fully unsupervised: the Generator is never fully connected to the input training data and thus needs to learn to generate data from noise without directly accessing the training data \cite{begusCiw} (for differences in performance between VAEs and GANs in the visual domain, see \cite{wu17}). Additionally, unsupervised ASR models increasingly include the GAN architecture \cite{baevski21}.

Finally, the output of the proposed technique \cite{begusZhou1} is directly interpretable time-series data. Our proposal requires no further processing of the outputs (such as PCA): the proposed technique results in time-series data from each convolutional layer that directly correspond to the waveform output in the final layer. This means that we can analyze outputs at the same time domain across the convolutional layers. Understanding encoding of intermediate representations in unsupervised models that operate with waveforms will be particularly important as ASR models increasingly operate with raw waveforms \cite{zeghidour18,baevski20}.

\section{Models}

\subsection{Model description}

The interpretation and visualization of individual layers is performed on the Generator network in two models: WaveGAN  \cite{donahue19} and ciwGAN \cite{begusCiw}. WaveGAN is a single-dimensional transformation of the Deep Convolutional GAN (DCGAN) architecture \cite{radford15} used for waveform data. Categorical InfoWaveGAN (CiwGAN) is an InfoGAN\cite{chen16} modification of WaveGAN that includes an additional ``Q-network" which forces the Generator to output informative data.

\begin{figure}
\centering
\includegraphics[width=.4\textwidth]{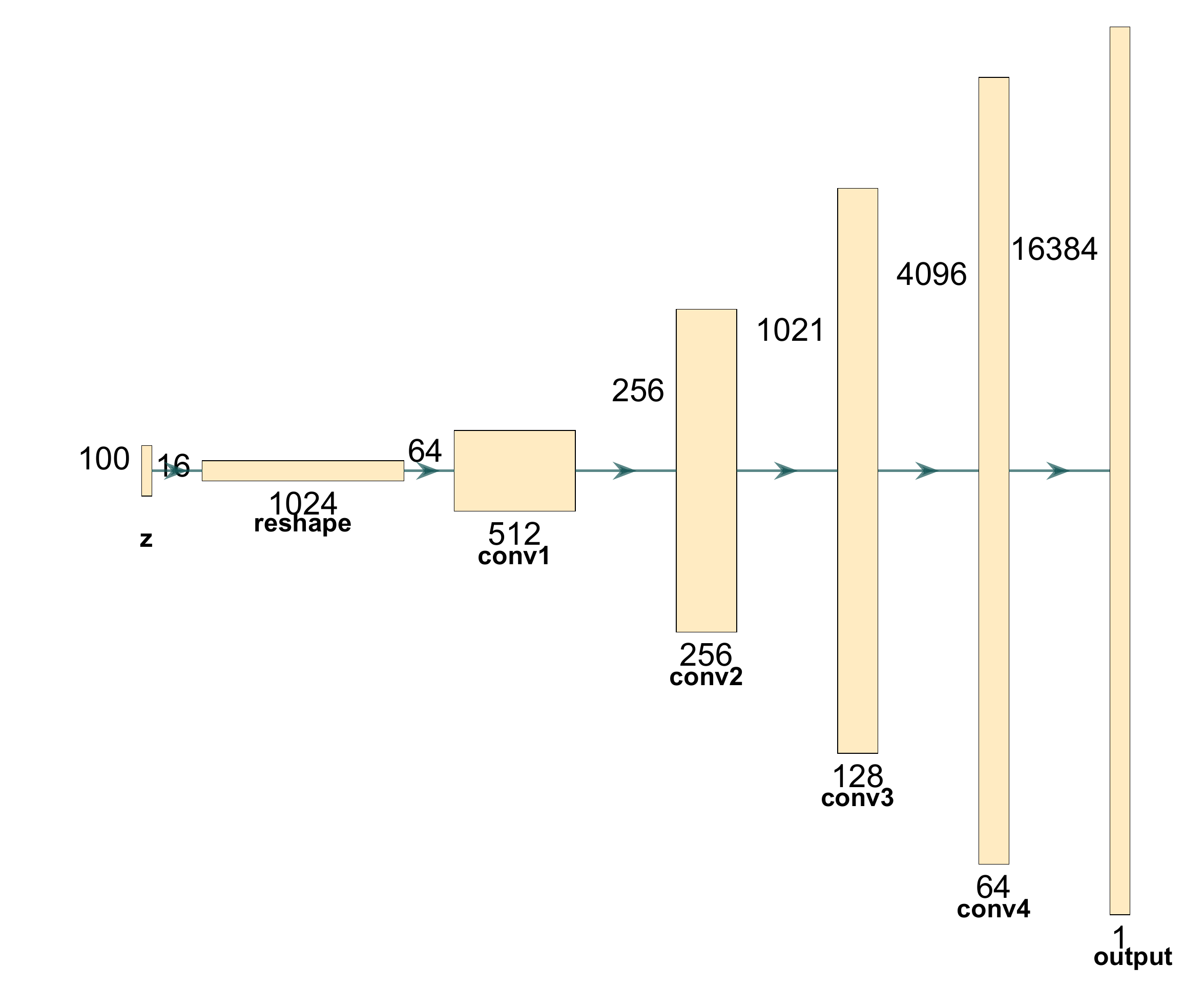}
\caption{The architecture of the Generator network with five one-dimensional convolutional layers as proposed in \cite{donahue19} and used for training in this paper. Filters are one-dimensional with the size of 25 \cite{donahue19}.}
\label{fig:nnTikz_gen}
\end{figure}

Both WaveGAN and ciwGAN contain a Generator and a Discriminator. The Generator takes 100 latent variables $z$ uniformly distributed in the interval $(-1,1)$ and transforms them into 16,384 data points constituting 1.024 s of audio file (sampled at 16 kHz) through five 1D convolutional layers. The dimensions of the five convolutions (four intermediate layers and the final output layer) are $512\times64\times1$, $256\times256\times1$, $128\times1024\times1$, and $64\times4096\times1$. The final layer (with tanh activation) has a dimension of $16384\times1\times1$.  All layers except for the last one are trained with ReLU activation. The dimensions are  summarized in Figure \ref{fig:nnTikz_gen}.

 The Discriminator network takes real and generated audio files (16,384 data points constituting audio file) and is trained using the Wasserstein loss formulation \cite{arjovsky17} with gradient penalty \cite{gulrajani17} (WGAN-GP). The Wasserstein distance between two distributions $P_X$ and $P_G$ is given by:
 
 \begin{equation}
 \mathcal{W}(P_X, P_G) = \sup_{\|f\|_{L} \leq 1} \mathbb{E}_{x\sim P_X}[f(x)] - \mathbb{E}_{x\sim P_G}[f(x)] \label{wasserstein}
 \end{equation}
 
 where $x$ describes datapoints sampled from each distribution and $\|f\|_L \leq 1$ is the family of 1-Lipshitz functions \cite{arjovsky17}. In WGAN-GP, we have the Discriminator take the place of $f$ in (\ref{wasserstein}), and use gradient penalty during training to ensure that the Discriminator remains 1-Lipschitz. We thus have the Generator and Discriminator participate in the following zero-sum game:
 
 \begin{equation}
 \min_{\theta_G} \max_{\theta_D} \left( \mathbb{E}_{x \sim P_X}[D(x)] - \mathbb{E}_{z \sim P_z}[D(G(z))]\right) \label{objective}
 \end{equation}
 
 where $\theta_G$ are the parameters of the Generator, $\theta_D$ the parameters of the Discriminator, $G$ the Generator, $D$ the discriminator, $P_x$ the training distribution, and $P_z$ the distribution of the latent space \cite{donahue19}. During training, the Generator and Discriminator take turns minimizing or maximizing the objective in (\ref{objective}), ideally reaching equilibrium when the approximated Wasserstein distance between the generated samples and real data is minimized.

\begin{figure}
\centering
\includegraphics[width=.47\textwidth]{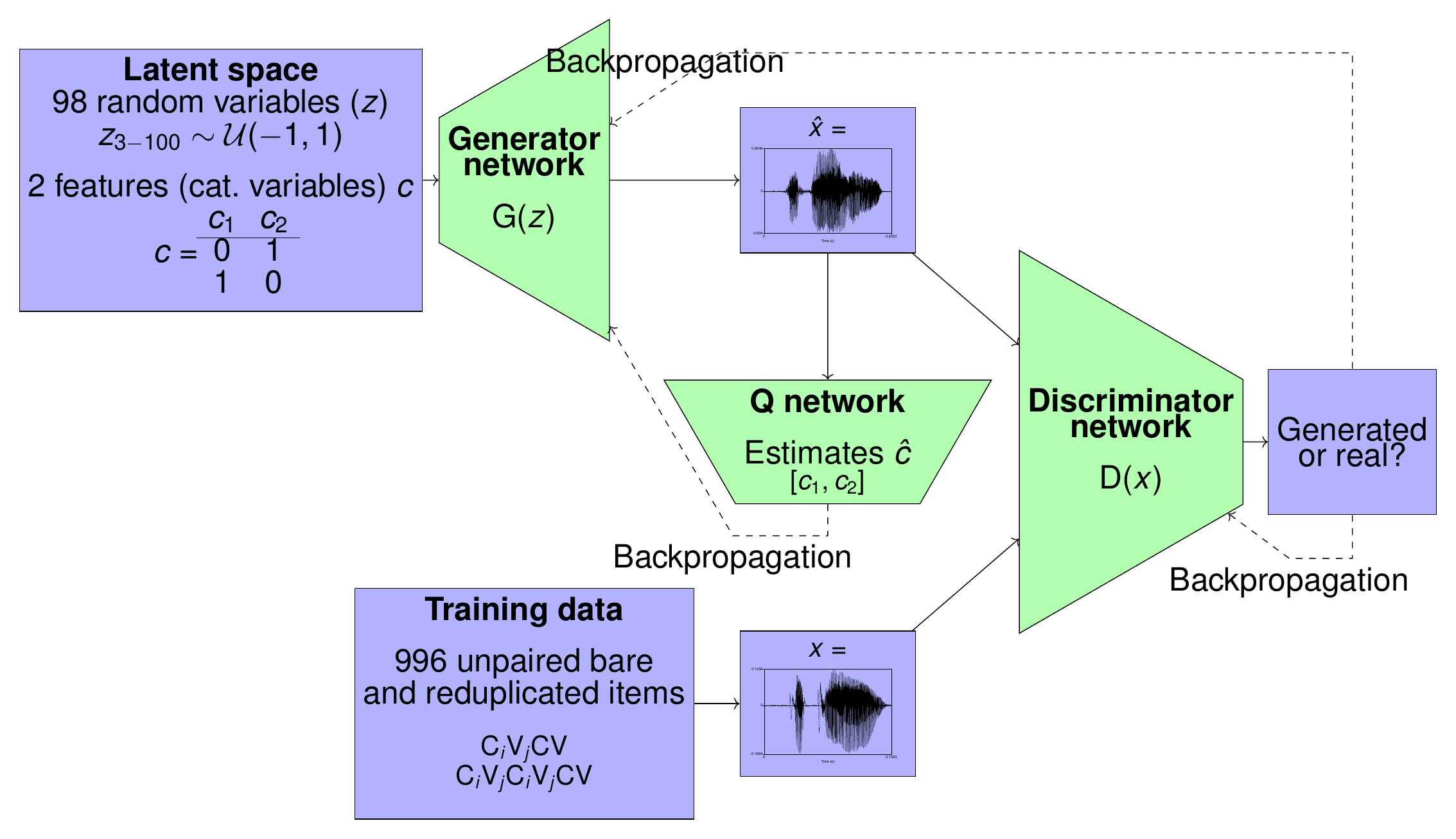}
\caption{The ciwGAN architecture as proposed in \cite{begusCiw} used for interpreting intermediate layers in Section \ref{model2}. Figure taken from  \cite{begusIdentity}.}
\label{fig:redgan}
\end{figure}

The ciwGAN architecture \cite{begusCiw} modifies the bare WaveGAN architecture by having the Generator take as input categorical code variables $c$ in addition to the latent variables $z$ and with the addition of a separate Q-network \cite{begusCiw} to estimate these categorical codes $c$. The Q-network and the Generator are trained to maximize the Q-network's success rates (the architecture is summarized in Figure \ref{fig:redgan}). The Q-network is in structure identical to the Discriminator except in its final layer, which is trained on estimating the Generator's latent code $c$ with a softmax function. In other words, the proposed architecture forces the Generator to output informative data. For example, when the ciwGAN network is trained on words from TIMIT, the most informative way to encode unique information (e.g.~a one-hot vector) into acoustic data is to associate each word with a unique latent code $c$. Lexical learning (associating acoustic lexical items with unique latent representation) thus emerges automatically from only the requirement that the Generator produce informative data in a completely unsupervised manner -- lexical items are never labeled or paired during training. Training thus results in a Generator that learns to output unique words for each latent code \cite{begusCiw}.

The addition of the Q-network modifies the training objective with an additional term of the cross-entropy between the predicted latent code $Q(G(z, c))$ and the true latent code $c$. \cite{begusCiw} adds this additional term to (\ref{objective}):

\begin{align}
\min_{\theta_G, \theta_Q} \max_{\theta_D} (&\mathbb{E}_{x \sim P_X}[D(x)] - \mathbb{E}_{z \sim P_z}[D(G(z))] \nonumber\\& - \lambda \mathbb{E}_{c \sim P_c, z \sim P_z}[\log Q(G(z, c))]) \label{qobjective}
\end{align}

where $\theta_Q$ are the parameters of the Q-network, $P_c$ is the distribution of the latent codes, and $\lambda$ is a tunable hyperparameter. The new cross-entropy term acts as a lower bound on the mutual information between the latent code and generated outputs, ensuring that the Generator uses informative latent codes in addition to generating realistic data.

\subsection{Model 1: Bare WaveGAN trained on a simple conditional distribution}
\label{bareganonan}

\subsubsection{Generator trained on \#TV and \#sTV sequences}

First, we analyze how the three basic acoustic properties of speech are encoded in CNNs: periodic vibration corresponding to vowels, aperiodic noise corresponding to fricatives (such as [s]), and silence corresponding to the closure part of stop consonants. For this purpose, we perform an analysis on the pretrained Generator network from \cite{begus19} on sliced sequences of the structure \#sTV and \#TV from TIMIT \cite{timit} (where T = /p, t, k/ V = vowel, \# = word edge). Altogether 5,463 data points from TIMIT were used for training: 4,930 sequences of the structure \#TV (such as [\textipa{"p\super h\ae}]) and 533 of the structure \#sTV (e.g.~[\textipa{"sp\ae}]). We used simplified training materials to facilitate interpretation of intermediate layers, but the visualization technique proposed here is scalable to more complex training data too. The network in \cite{begus19} was trained  for 12,255 steps (approximately 716 epochs). At this point, the network  not only learns to output  speech-like sequences  (\#TV and \#sTV) that resemble training data and are acoustically analyzable, but also  learns the simple conditional distribution in which aspiration is shortened if [s] is present in the output ([\textipa{"p\super h\ae}] vs.~[\textipa{"sp\ae}]) \cite{begus19}.

\subsubsection{Finding linguistically meaningful units} In \cite{begus19}, a technique is proposed that identifies those latent variables from $z$ that correspond to some meaningful linguistic representation in the output, such as presence of [s]. The technique includes  training the network, generating data, and annotating them for presence of any acoustic or higher level phonological property (in our case, presence of frication noise of [s] or presence of reduplication). In \cite{begus19}, annotation is performed manually, but automated annotations can be employed as well. The data with presence or absence of an acoustic property as the dependent variable and the 100 latent variables as predictors is then fit to regression models which identify those variables in the latent space $z$ that most strongly correspond to the presence of the phonetic or phonological property in question. Based on results from the regression model, \cite{begus19} argues that the Generator learns to represent the presence of [s] with a subset of latent variables $z$. Crucially, it is shown that manipulating the variables chosen with the regression technique results in an almost one-to-one mapping between \emph{individual} latent variables and the presence of [s]. Several generative tests are performed to confirm the link between individual latent variables and the presence of some linguistically meaningful unit. \cite{begus19} proposes that by manipulating individual variables to levels well outside of training range (i.e.~well outside the interval $(-1,1)$, which are called marginal levels henceforth) to values such as $\pm15$, we can force [s] to surface (or not surface) in the output at near categorical levels \cite{begus19,begusCiw}. The value $\pm$15 was chosen because the output does not change substantially with values higher than $\pm$15. The range of $z$-values that yield informative outputs during linear interpolation differ across models and likely depends on the number of training data points and diversity of the data (as differences in \cite{begusIdentity,begusLocal} suggest). 

Manipulating individual latent variables to marginal values well outside of the training range to create a high occurrence of a desired linguistic unit \cite{begus19} is a crucial concept used in this paper.  This technique reveals that the Generator learns to use the latent space as a discretized representation of linguistically meaningful units. For example, using regression techniques, 7 variables $z_i$  out of the 100 in the latent space are identified in \cite{begus19} that strongly correspond to presence of [s] in the output. These variables are learned during training and will vary with different training trajectories for the same model.  The eleventh variable $z_{11}$ is one such variable that strongly corresponds to presence of [s].  By setting $z_{11}$ to $-1$ (within the training range), we get a modest increase of [s]-containing sequences in the output. By setting it to $-15$, 87\% of outputs contain an [s]; by setting it to $-25$, there are 96\% such outputs \cite{begus19}.

Begu\v{s} \cite{begus19} shows that in the model trained on \#TV and \#sTV sequences, linearly interpolating $z_{11}$ from marginal values results in a gradual reduction of frication noise in the output until [s] ceases from the output; the frication noise of [s] appears to be directly causally connected with $z_{11}$. To linearly interpolate a variable from marginal values, we generate a set of linearly spaced points along the interval between the marginal values and set the variable to each of those values. Figure \ref{fig:3dscatter} (bottom right) shows how linearly interpolating $z_{11}$ from 5 (corresponding to absence of [s] in the output) to $-15$ (corresponding to presence of [s] in the output) results in the gradual appearance and then increase of frication noise (corresponding to [s]) in the generated output. In \cite{begus19},  it is shown that direct correlations between single latent variables and the amplitude of  frication noise of [s] in the output operate across generated samples and persists even when the amplitude is measured proportionally  to the vocalic amplitude. In sum, there is a causal relationship between individual latent variables identified with  the proposed technique \cite{begus19} and linguistically meaningful properties of the output.

\subsection{Model 2: Deeper WaveGAN trained on LibriSpeech}
\label{deepbareganonan}
In order to test how intermediate representations vary across model size and application and how the proposed technique scales up to larger models trained on larger corpora, we additionally train a deeper WaveGAN model \cite{donahue19}. For the purpose of this paper, we increase the depths of both the Generator and Discriminator networks from 4 intermediate convolutional layers to 9 intermediate layers. All other parts of the model architecture were unaltered. The exact dimensionalities of each layer are described in Supp.~Materials Table 1.

We train the model on 559,992 tokens of 508 words sliced from the LibriSpeech  corpus \cite{librispeech} for 34,577 steps, after which mode collapse was observed. These words were chosen by discarding the 78 most common words that appeared disproportionately more frequently in LibriSpeech train-clean-360 (ranging from 5,290 to 224,173 tokens per word), and arbitrarily choosing the next 508 most frequent words (571 to 5,113 tokens per word).

\subsection{Model 3: CiwGAN trained on an identity-based pattern}
\label{ciwganonan}

The conditional allophonic distribution described in \ref{bareganonan} is computationally among the simplest processes in human languages. To test whether the technique for interpretation of intermediate layers extends to computationally more complex processes in language, we apply the technique to a pretrained ciwGAN model on an identity-based pattern (copying) called reduplication (in \cite{begusIdentity}). 

The advantage of the ciwGAN architecture is that learning of linguistically meaningful units emerges from the requirement that the Generator outputs informative data. To test how learning of a highly complex process such as reduplication self-emerges in this architecture, \cite{begusIdentity} trains the ciwGAN network  with one-hot latent code of length 2 on 996 bare and reduplicated items (e.g.~[\textipa{"p\super hAli}] and [\textipa{p2"p\super hAli}]). The bare and reduplicated forms are never paired in the training data and are presented randomly. The model is trained for 15,920 steps (or approximately 5,114 epochs). The Generator learns to associate the latent code with reduplication: when latent code (one-hot vector with two levels) is set to marginal levels of 5 [5, 0], the Generator outputs 98\% unreduplicated bare forms; when it is set to [0, 5], it outputs 87\% reduplicated forms \cite{begusIdentity}. When the values are linearly interpolated, the Generator gradually turns a bare unreduplicated form into a reduplicated form (e.g.~from [\textipa{"p\super hi\*ru}] to [\textipa{p@"p\super hi\*ru}] \cite{begusIdentity}) in approximately 50\% of outputs that undergo the change from bare to reduplicated (25\% of total outputs). Figure \ref{fig:3dscatterCiw} (bottom right) shows how manipulating categorical latent variable $c_2$ results in the gradual  appearance of a reduplicated syllable in the output.  The network also learns to extend the learned pattern to unobserved data and reduplicates forms with initial consonants that were withheld from training \cite{begusIdentity}. For example, by simultaneously forcing reduplication and [s] in the output (setting the latent variables to marginal levels beyond training range), the network outputs [\textipa{s@"siji}], although [\textipa{s@"siji}] and all [s]-containing reduplicated forms were withheld from training data (the network only sees unreduplicated [s]-initial words such as [\textipa{"siji}]). These results \cite{begusIdentity} strongly suggest that the Generator learns to represent a linguistically meaningful and computationally highly complex process (reduplication or copying) with the latent codes in a fully unsupervised manner.

In \cite{begus19,begusCiw,begusIdentity,begusLocal}, we only analyze and interpret the endpoints of these models: the latent variables and the generated outputs. Here, we propose that intermediate convolutional layers can be interpreted using this technique as well. 

\section{Interpretation}

We propose that learned representations in the intermediate layers can be evaluated by combining two techniques: (i) averaging across feature maps in each layer after ReLU activation (as in \cite{begusZhou1}; Sections \ref{sub1} through \ref{model2}) and (ii) manipulating individual $z$ variables to marginal values well outside the training range (as in \cite{begus19}; Section \ref{2nd}). 

Averaging across feature maps yields interpretable time-series data at each convolutional layer that shows how features are encoded in each layer \cite{begusZhou1}. In short, for each convolutional layer $C \in \{\text{Conv1, Conv2, Conv3, Conv4}\}$, we perform the averaging operation (from \cite{begusZhou1})

\begin{equation}
    \frac{1}{\|C\|} \sum_{i=1}^{\|C\|} C_i\label{averaging}
\end{equation}

where $C_i$ is the $i$th feature map of layer $C$ and $\|C\|$ is the total number of feature maps in $C$. This yields a time series that summarizes the information encoded at each layer. 

To evaluate the causal relationship between individual latent variables and the convolutional layers, the $z$ variables can be linearly interpolated from marginal endpoints outside of the training range. The proposed technique reveals which features in the intermediate layers get activated when manipulating individual latent variables $z$ and which linguistically meaningful variables (such as duration, F0, intensity, or formant structure) get encoded at which layers. For a particular dimension of the latent noise $z$, we interpolate linearly between two extreme values and observe changes in the intermediate representations. For example, to test the 11th dimension of the latent space $z_{11}$ as it changes from $-5$ to $5$, we freeze the rest of the latent space and vary $z_{11}$ through the values $-5, -4.5,  -3, \dots, 3, 4.5, 5$ (linear interpolation with a constant interval of 0.5), observing the Generator's outputs and intermediate representations at every step.

This approach also allows us to follow how linear interpolation of individual latent variables $z$ (such as $z_{11}$) that correspond to some meaningful linguistic unit (such as presence of [s] or reduplication) affect individual feature maps in each convolutional layer (Section \ref{individual}). 

\subsection{Model 1: WaveGAN \cite{donahue19}}
\label{sub1}

\begin{figure}
\centering
\includegraphics[width=.47\textwidth]{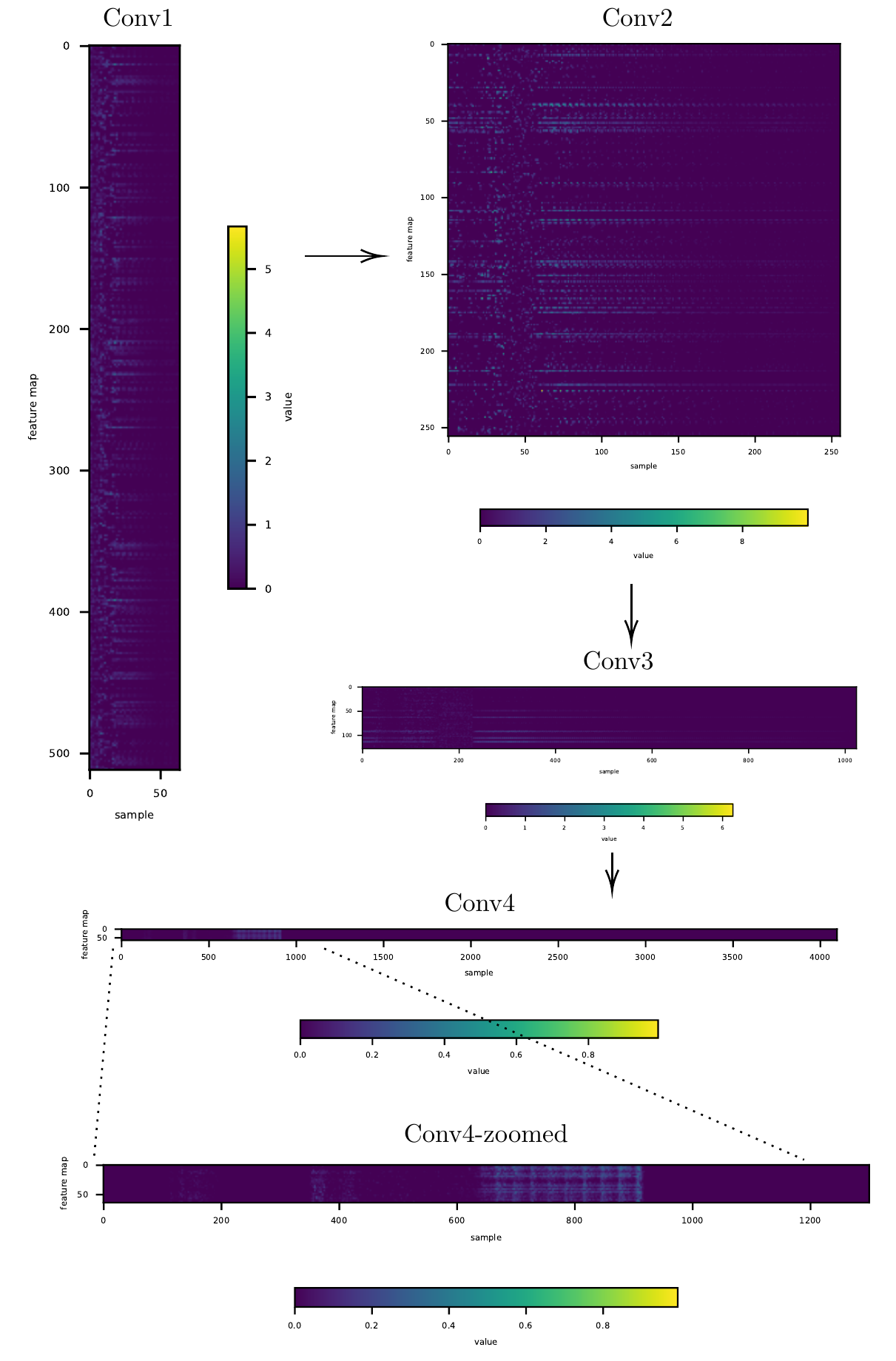}
\caption{Values of feature maps (concatenated on the y-axis) after ReLU activation in  four convolutional layers  for a uniformly distributed $z$-vector limited to the training interval $(-1,1)$. The visualizations illustrate how activations in the previous layers result in a clearly analyzable periodic vocalic structure in the fourth convolutional layer  (Conv4 on the zoomed-in graph) that in turn results in a periodic vocalic vibration in the output.  }
\label{fig:featureMaps}
\end{figure}

Figure \ref{fig:featureMaps} plots  values of each feature  map (concatenated along the y-axis) for a $z$ that is uniformly distributed on the training interval $(-1,1)$ across all variables. The visualization illustrates the structure of the Generator. At the fourth convolutional layer, a clear periodic structure of the vocalic part is visible. The most common technique of visualizing CNNs --- a simple concatenation of feature maps  ---  does not provide the most interpretable results in speech  beyond these basic observations. 

\begin{figure}
\centering
\includegraphics[width=.4\textwidth]{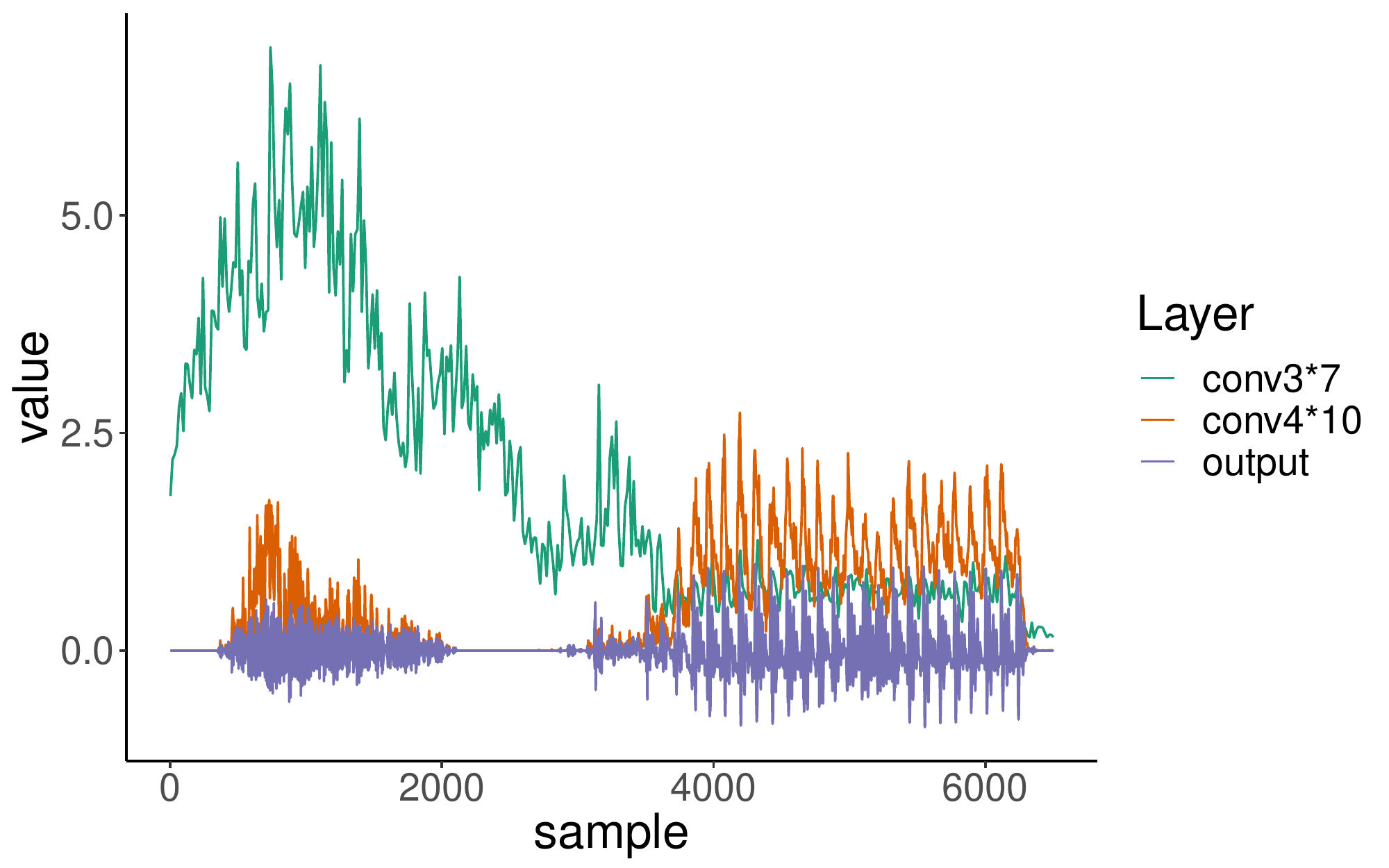}
\includegraphics[width=.4\textwidth]{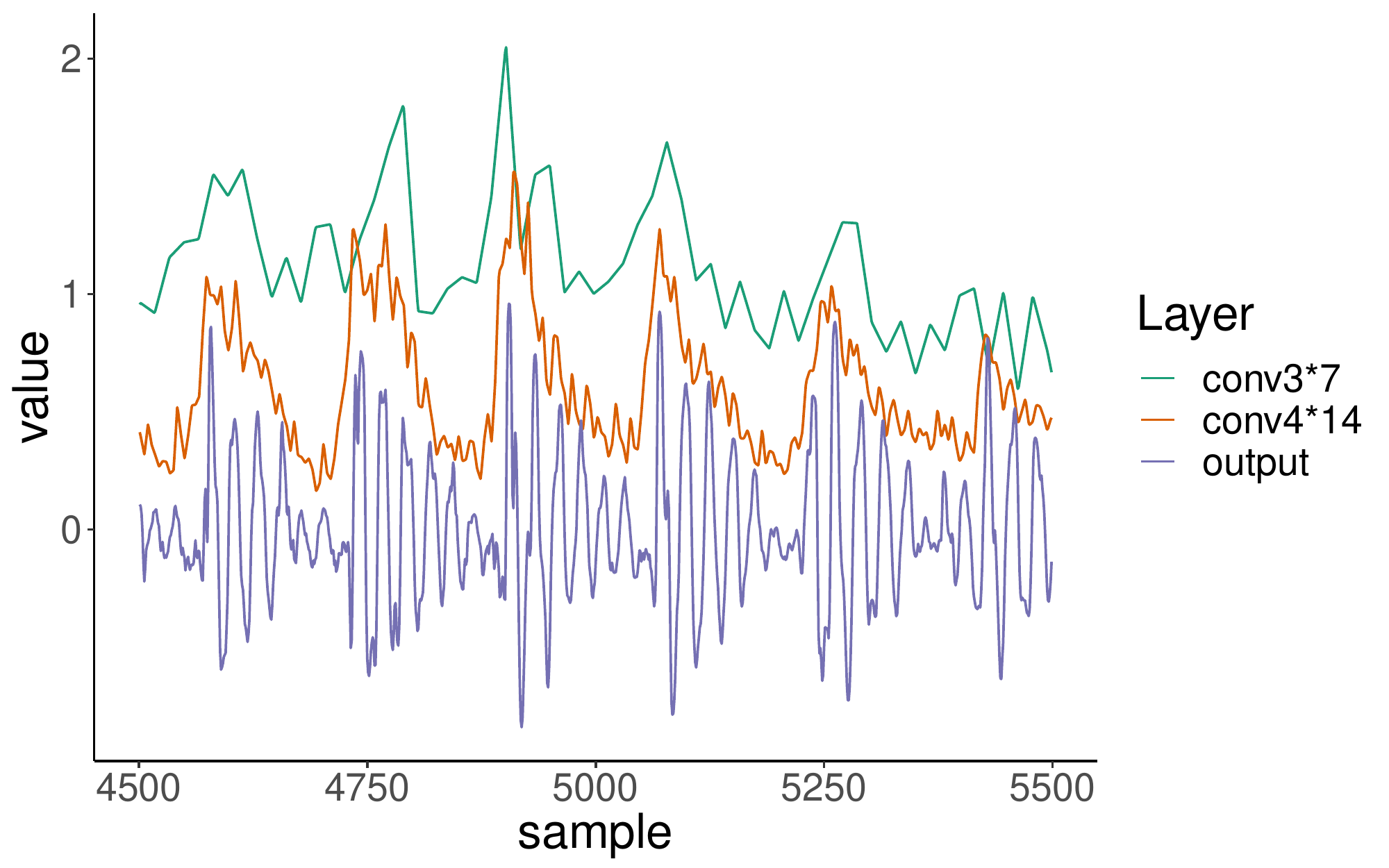}
\includegraphics[width=.4\textwidth]{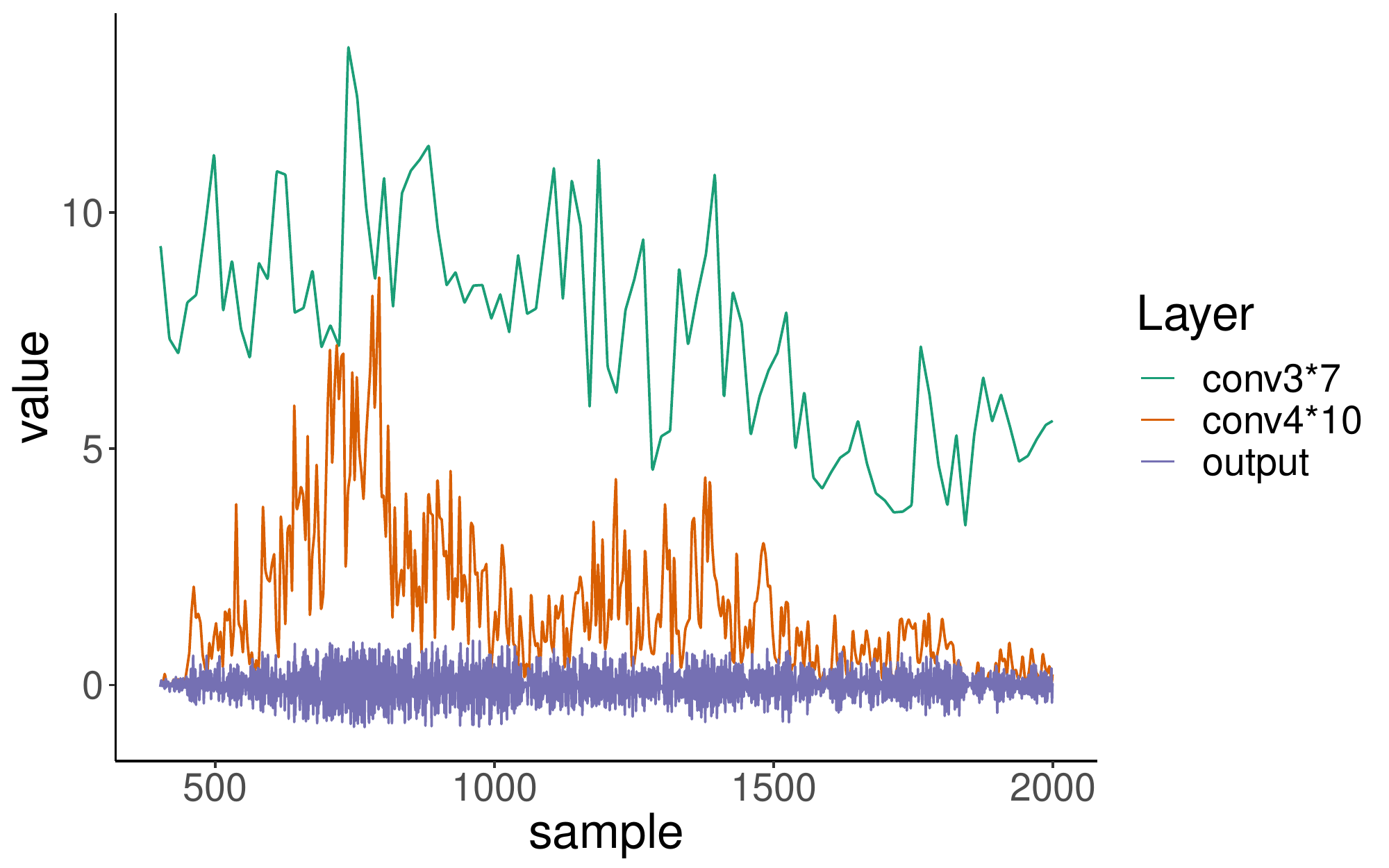}
\caption{All feature maps averaged after ReLU activation  after the third convolutional layer (conv3; green), fourth convolutional layer  (conv4; dark orange) and the generated output (output; purple). (top) A generated output when $z_{11}=-5$ featuring a period of frication [s], a period of silence (of a stop consonant), and a vocalic period. To overlay the two convolutional layers on top of the output, they are multiplied by 7 and 10, respectively. (middle) A period of vocalic periodic vibration with the same latent space values as above, but $z_{11}$ set at -1 and conv3 and conv4 multiplied by 7 and 14, respectively, to overlay the convolutional layers on top of the output. (bottom) A period of frication (in [s]) with the same latent space values as above, but $z_{11}$ set at -11 and conv3 and conv4 multiplied by 7 and 10, respectively, to overlay the convolutional layers on top of the output. }
\label{fig:out}
\end{figure}

Averaging across all feature maps as in equation \ref{averaging} results in highly interpretable time-series data. Figure \ref{fig:out} plots the third (Conv3) and fourth (Conv4) convolutional layers, averaged across all feature maps after ReLU activation  along with the corresponding waveform output that can be transcribed as involving a fricative [s], a stop, and a vowel (\#sTV). Overlaying the last two convolutional layers with the generated output reveals that the fourth convolutional layer includes information for all three major acoustic properties of the output: we observe a period of aperiodic vibration corresponding to the frication noise (in [s]), a period of silence corresponding to the closure portion of the consonant (T) and a clear periodic vibration corresponding to the vowel (V). The timing of these constituents in Conv4 aligns almost perfectly with the generated output.

The fourth layer (Conv4) carries both the fundamental frequency (F0) and formant structure information in the vocalic part of the input. Figure \ref{fig:out} (middle) clearly shows that the averaged fourth convolutional layer after ReLU contains periodic vibration with the fundamental frequency that matches the output as well as higher-frequency vibration that  corresponds to the formant structure in the output. Amplitude/intensity information also appears to be encoded in the fourth layer --- Conv4 closely traces the actual output in the final layer.

\begin{figure}
\centering
\includegraphics[width=.4\textwidth]{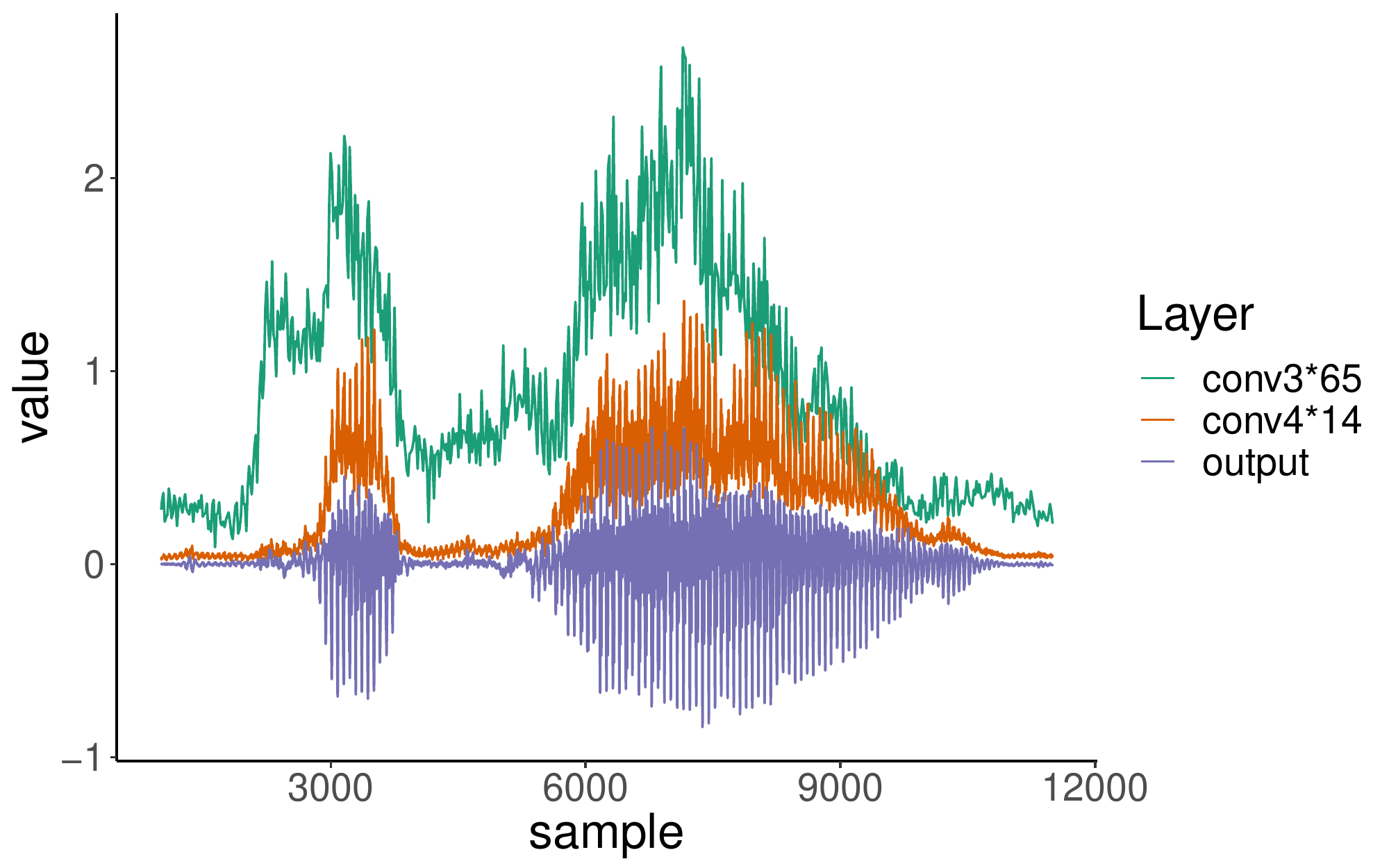}
\includegraphics[width=.4\textwidth]{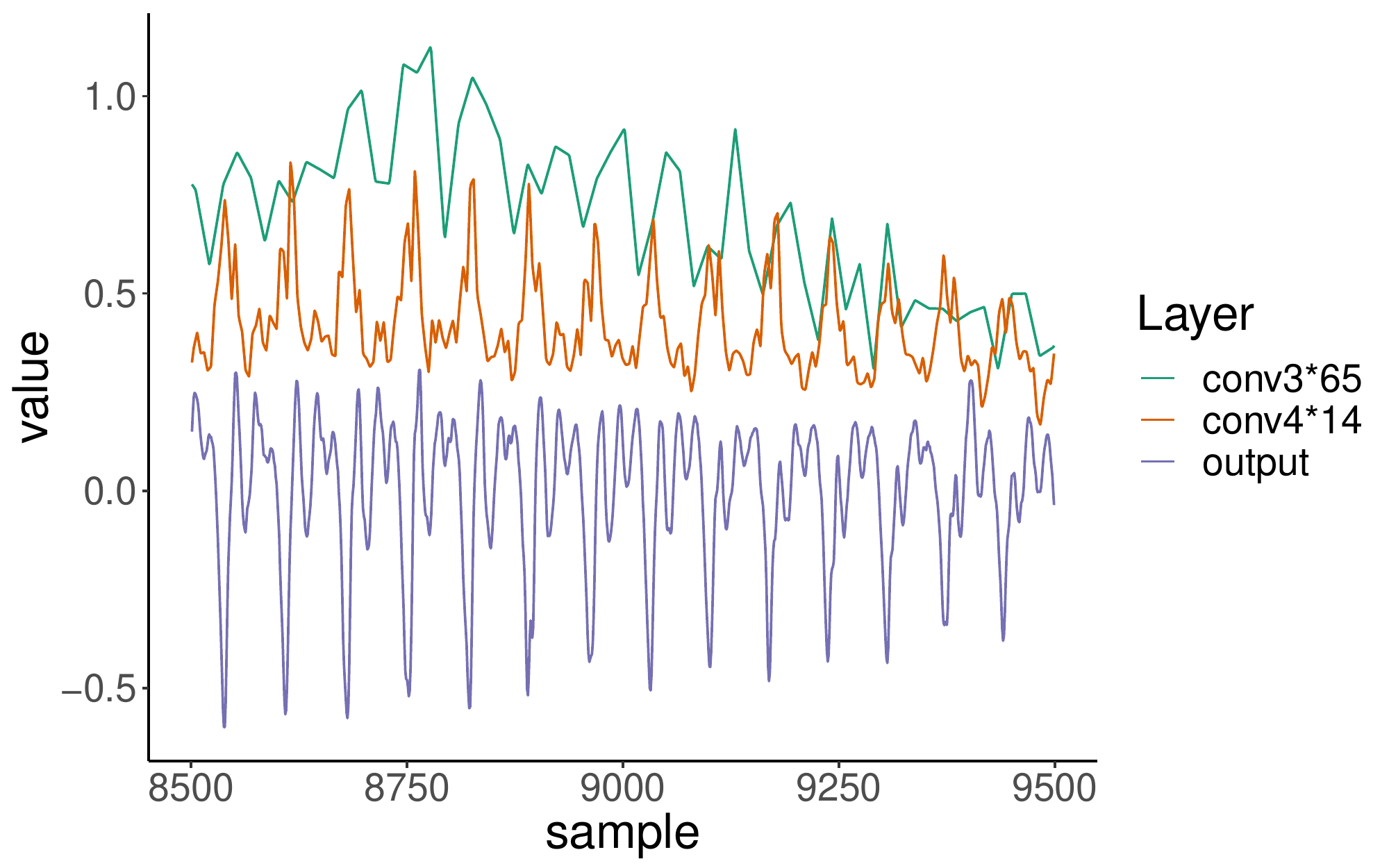}

\caption{All feature maps averaged after ReLU activation after the third convolutional layer (conv3; green), fourth convolutional layer  (conv4; dark orange) and the generated output (output; purple). (top) A generated output when c$_{2}=1$. To overlay the two convolutional layers on top of the output, they are multiplied by 65 and 14, respectively. (bottom) Zoomed-in enerated output when c$_{2}=1$.}
\label{fig:redout}
\end{figure}

To quantify these observations, we randomly generate 25 outputs from the bare WaveGAN model trained on \#TV and \#sTV sequences \cite{begus19} and convert outputs from intermediate layers to waveforms ready for acoustic analysis.\footnote{As the intermediate layers are all positive, we clip all values greater than 1 to be equal to 1 in the waveform outputs. We then treat the signal as a float32 signal and convert it to a .wav file.  We also upsample the intermediate layers to 16 kHz sampling with linear interpolation.} We manually annotate the vocalic period in the final output and perform acoustic analysis of the outputs in the third and fourth convolutional layers (Conv3 and Conv4).

\subsubsection{Duration}
\label{durationwav1}

We manually annotate periodic vibration in the fourth convolutional layer and compare vowel durations of the 25 generated outputs between the final output and the fourth convolutional layer. The vocalic durations are easily identifiable in Conv4 and nearly identical to the vocalic duration in the final output. Durations from the two layers fit to a linear model reveal a high degree of correlation ($\beta = 0.96,t=  30.31,p <0.0001$) with adjusted $R^2=0.97$ (Supp.~Materials Figure 17). 
In the averaged Conv3-output, the difference between the periodic vibration characteristic of vowels and other acoustic properties, such as silence (characteristic of stops) or frication noise (characteristic of fricatives and aspiration), are not clearly visible (see Figure \ref{fig:3dscatter} and Supp.~Materials Figure 20). 

Based on these results, we can conclude that vocalic  duration and periods of silence corresponding to stop closure are most strongly encoded in the fourth convolutional layer (Conv4) in the model trained on \#TV and \#sTV sequences.

\subsubsection{F0}

\begin{figure}
\centering
\includegraphics[width=.48\textwidth]{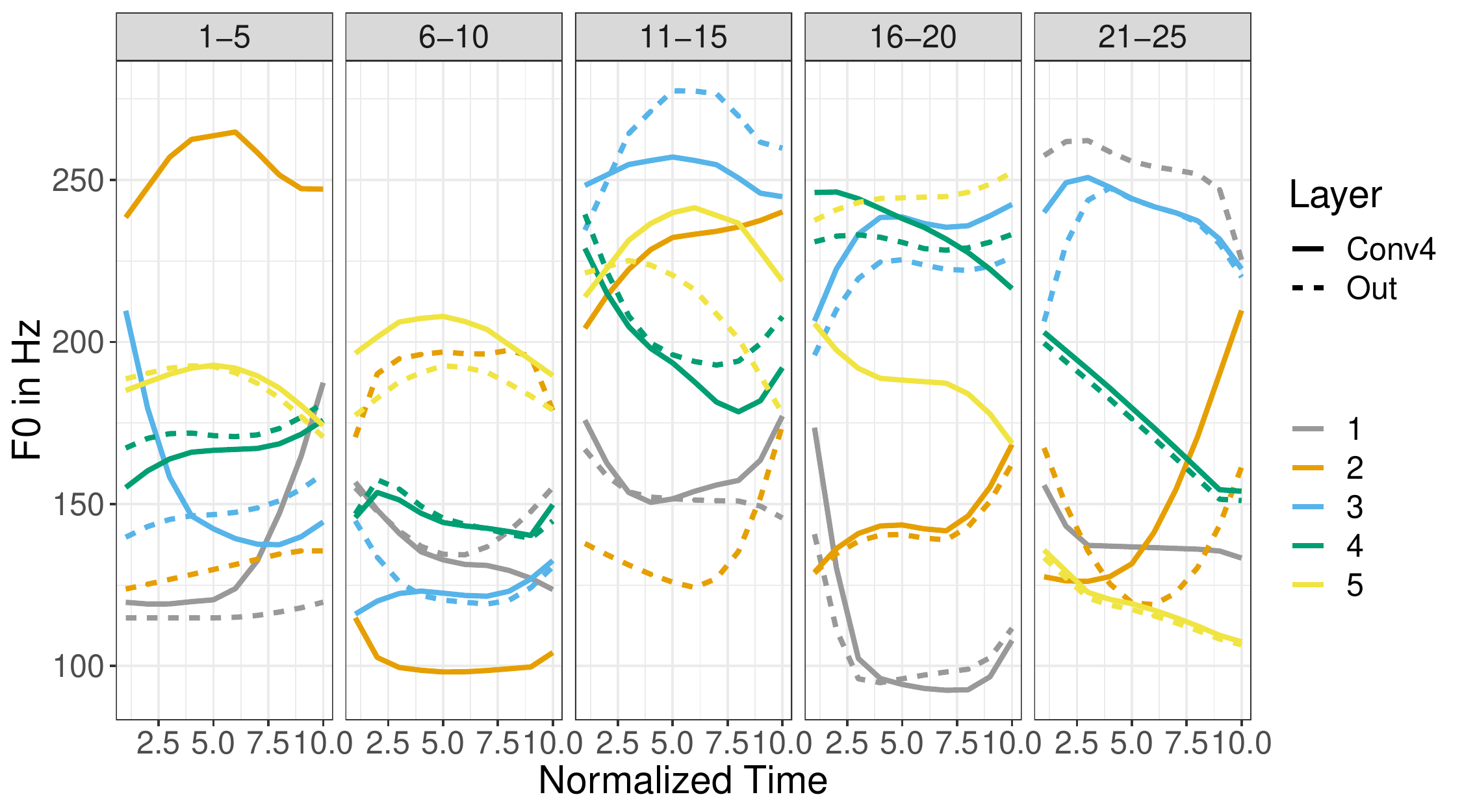}
\\
\includegraphics[width=.48\textwidth]{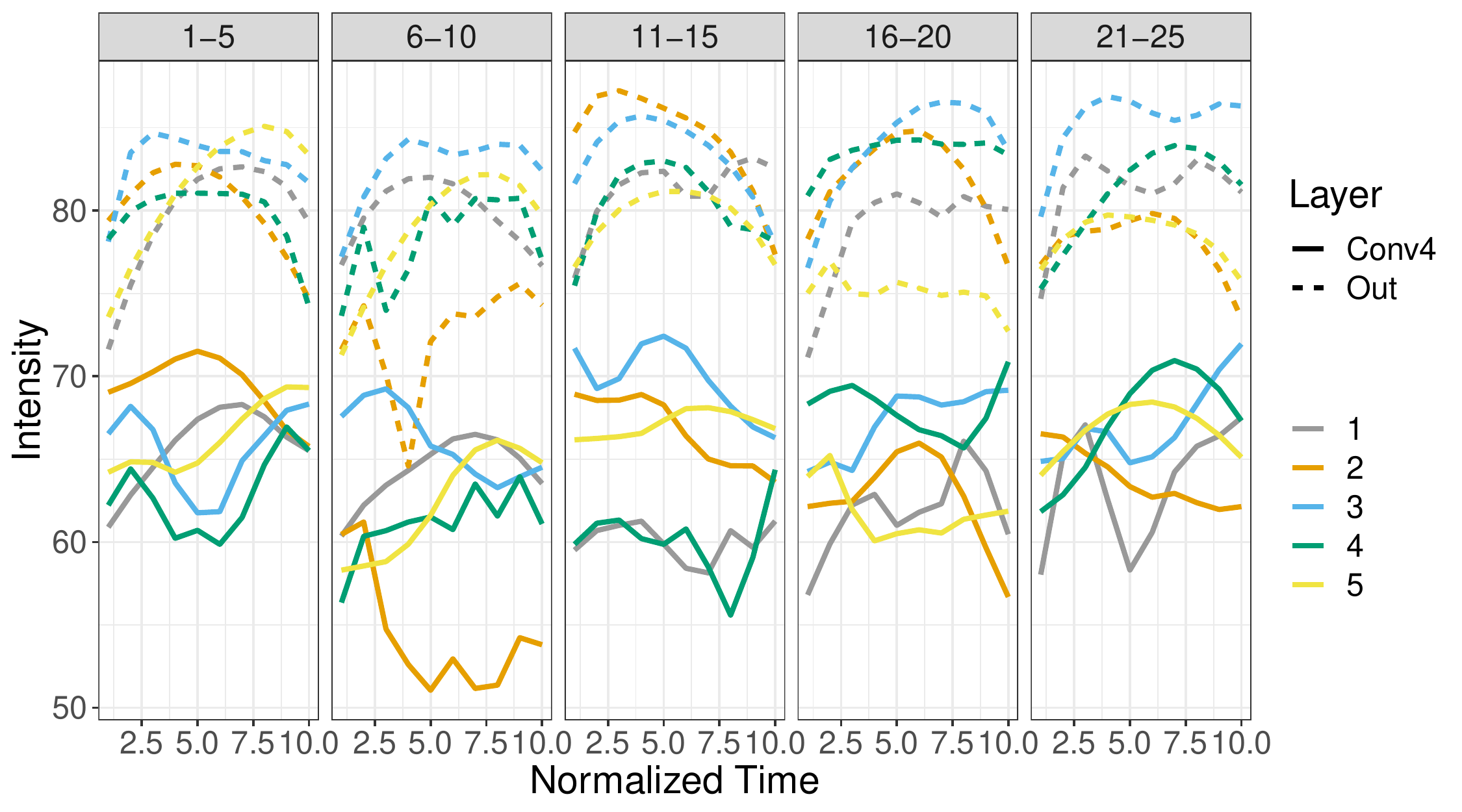}

\caption{(top) F0 values in normalized time (10 intervals) in  25 randomly generated outputs for the final output (Out) and fourth convolutional layer (Conv4), grouped in five bins (1-5) for presentational purposes. The values were extracted using the Praat software \cite{boersma15} with a script by Xu \cite{xu13}. The window for F0 range was set to 60-300Hz for the analysis. (bottom) Intensity values (in dB) in normalized time (10 intervals) in  25 randomly generated outputs for the final output (Out) and fourth convolutional layer (Conv4). The values were obtained as described for F0 (the minimum pitch for  intensity is 100  Hz).}
\label{fig:F0}
\end{figure}

To test how the Generator encodes the fundamental frequency (F0), we extract F0 values from the manually annotated vocalic period in the 25 randomly generated outputs.\footnote{For the purpose of analyzing F0 and intensity, we use  annotations of the vocalic period from the final output (Out) also for the analysis of F0 and intensity in the third and fourth convolutional layers (Conv3 and Conv4).}
The Conv4 outputs are noisy and limited to positive values, which is why extraction of F0 can be challenging. F0 values are extracted using Praat script by  Xu \cite{xu13} with the range of F0 set to 60--300 Hz for the analysis. Figure \ref{fig:F0} shows the 50 extracted values (25 for each layer). Several F0 trajectories are almost identical between the final layer and Conv4. A correlation test of concatenated values between the two layers (Conv4 and output) reveals a substantial correlation with $r=0.53$ (Pearson's product-moment correlation henceforth, marked with $r$). The correlation is calculated on all outputs together  with no levels for individual outputs (here and in the following cases).

Figure \ref{fig:out} suggests that F0 is likely also encoded in Conv3. The Conv3 layer shows peaks that correspond to vocalic periodic vibration.  However, with the relatively weak signal, F0 contours are difficult to extract from the Conv3 of a model that is trained with relatively few steps. For further discussion on F0, see Section \ref{f0red}.

\subsubsection{Intensity}
\label{stvintensity}

To test whether and how intensity is encoded in Conv4 (as observed in the qualitative analysis in Figure \ref{fig:out}), we extract intensity values from annotated vocalic periods (using the the script by Xu \cite{xu13} in Praat \cite{boersma15} with 100 Hz minimum pitch and annotated in the final output layer). Figure \ref{fig:F0} illustrates that the intensity values of Conv4 are lower compared to the final output, but there is a substantial correlation  between  concatenated values of intensity  in the two layers:  $r=0.62$. 
Lower absolute values of the intensity levels are expected as the Conv4 layer only includes positive values and there is no reason for the network to match intensity values in absolute terms across the layers. 

We also correlate intensity levels between the third convolutional layer (Conv3) and the final output. Because vocalic period is not clearly encoded in Conv3, we use annotations of the vocalic period from the final output. There is a modest correlation in intensity values between Conv3 ant the final output:  $r=0.39$. 
Figure \ref{fig:3dscatter} also suggests that intensity (or amplitude envelope) is encoded in Conv4, Conv3, and perhaps even in Conv2 when individual latent variables are manipulated to marginal values.

\subsection{Model 2: Deeper WaveGAN \cite{donahue19}}

To test how the proposed technique scales up to larger models and how linguistically meaningful properties are encoded across the convolutional layers in deeper models, we focus on F0 encoding in the Deeper WaveGAN model trained on LibriSpeech. 

To test encoding of F0, we manually annotate 25 randomly generated outputs for periodic vocalic vibration in the final layer. F0 values in both the final layer as well as in higher convolutional layers  are extracted based on the annotations from the final layer. Ten F0 values are measured for each instance of vocalic vibration. We extract F0 values from Conv6, Conv7, Conv8, and Conv9  and compare them to extracted values from the final layer (Out).

Pearson's product-moment correlation reveals a high degree of correlation between Conv9 and output: $r=0.85$. This value is even higher than the correlation in the 5-layer WaveGAN, despite it being trained on substantially more data points. With each subsequent layer, the correlation gets smaller ($r=0.73$ for Conv8, $r=0.25$ for Conv7, and $r=0.15$ for Conv6). Supp.~Materials Figure 18 illustrates a high correlation in F0 between Conv9 and output and a substantially lower correlation between Conv6 and output (plots for Conv7 and Conv 8 are in Figure 19 in Supp.~Materials). We observe a similar trend in the deeper model as in the original WaveGAN: there is a steep drop in  correlation estimates  in F0 encoding between Conv9 and Conv8 vs.~Conv7 and Conv6. If we consider the  layers in the 10-layer model as a doubled 5-layer model, Conv9 and Conv8 together correspond to Conv4, while Conv7 and Conv6 correspond to Conv3. In the 5-layer model, there was a substantial drop in correlation estimates between Conv4 and Conv3, similarly to what we observe in the 10-layer model.

\subsection{Model 3: CiwGAN \cite{begusCiw}}
\label{model2}

Visualization of intermediate layers in Figure \ref{fig:3dscatterCiw} suggests that lower-frequency properties such as acoustic envelope are encoded in earlier convolutional layers and that properties with frequencies higher than acoustic envelope (such as F0 or formant structure) get added on top of the envelope outline in the later layers. To quantify this observation, we perform correlation analysis on the ciwGAN model. The ciwGAN model captures longer time frames of periodic vibration with more variable acoustic envelopes compared to the WaveGAN model because the training data involve words longer than a single syllable. The ciwGAN model also contains a more complex linguistic process --- reduplication (see Section \ref{intro} for a discussion on complexity).
We analyze the encoding of acoustic envelope (intensity) and F0 through all convolutional layers (Conv1-4 and the final output). We show that intensity is encoded in both the earlier layers and well into the deepest layers with high correlation estimates, whereas F0 gradually appears in later layers. Formant structure is encoded only in the final layer (it cannot even be tested in earlier layers).

We generate 30 random outputs, 15 each for the two values of the code variables ([0, 1] and [1, 0]). We extract F0 and intensity values over the entire periodic vibration of an output (all voiced sounds) based on authors' manual annotations. For example, in an output transcribed by the authors as [\textipa{"bAli}], the F0 and intensity values are extracted from all sounds, because they are all voiced.\footnote{For reduplicated outputs interrupted by a stop, we extract the values separately for  each periodic vibration, which totals in 38 analyzed periods from 30 outputs. }

\subsubsection{Intensity}

Intensity appears to be strongly encoded at all convolutional layers.  Contrary to the analysis in Section \ref{stvintensity}, intensity values in this model span not only a single vowel but often multiple vowels and voiced consonants (both sonorants and stops). Correlation between the concatenated final output values and averaged Conv4 values are high: $r=0.82$ (Figure 
21 in Supp.~Materials).
The correlation between the output and averaged values from the third, second, and first convolutional layers is slightly smaller, but nevertheless relatively high: $r= 0.72$ for Conv3, $r= 0.63$ for Conv2, and $r= 0.44$ for Conv1.

\subsubsection{F0}
\label{f0red}

 Outputs from  the ciwGAN model suggest that F0 is already encoded in the fourth convolutional layer, similarly to what is suggested in the bare WaveGAN model. The extracted F0 values often suffer from doubling and halfing errors, but there is still a correlation between F0 in the output and in the fourth convolutional layer (Conv4): $r=0.55$. 
 
\begin{figure*}
\centering
\includegraphics[width=.85\textwidth]{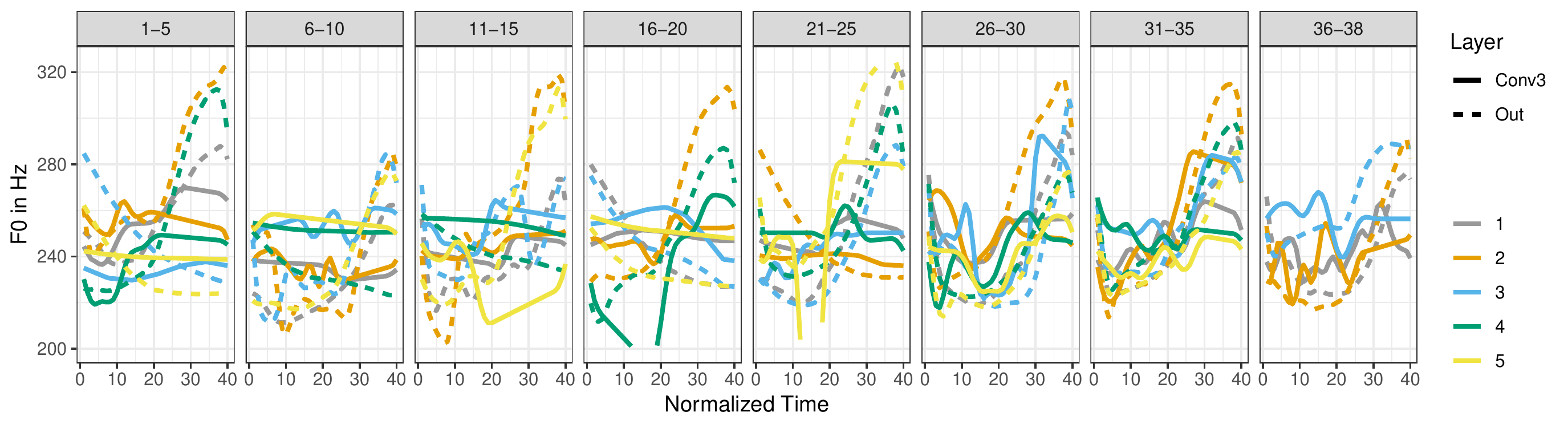}

\caption{F0 values in normalized time (40 intervals) in  30 randomly generated outputs (15 for each code; 38 vocalic periods total) for the final output (Out) and third convolutional layer (Conv3), grouped in five bins (1-5) for presentational purposes. The values were extracted using the Praat software \cite{boersma15} with a script by Xu \cite{xu13}. The window for F0 range was set to 75-450 Hz for the analysis. Values below 250 Hz and above 325 Hz are excluded from the plot.}
\label{fig:redF0}
\end{figure*}

 The ciwGAN model also suggests that the F0 is at least partly encoded already in the third convolutional layer (Conv3), but not earlier than that. Figure \ref{fig:redF0} plots all extracted F0 values from the final output and the third convolutional layer. There is a moderate correlation in F0 between the averaged Conv3 layer and the final output ($r=0.40$). In earlier layers, correlation is very low: $r=0.10$ for Conv 2 and $r=-0.02$ for Conv1 (despite the window for F0 being lowered to 5-150 Hz). 

Each convolutional layer is limited in what acoustic information it can encode directly as raw time series data by the Nyquist frequency: the layer's dimensions need to be at least twice the frequency of the acoustic property that needs to be encoded. For example, convolutions higher than the third layer (Conv3) cannot encode F0 in a non-abstract way: with a dimension of only 256, its Nyquist frequency is only 128 Hz. It is of course possible that different F0 values and trajectories are encoded in an abstract reduced representation in higher convolutions as well as in the latent space, but they cannot be encoded directly with frequency encoding. In principle, acoustic properties could be encoded with a quotient of frequency. For example, F0 could be encoded with halved values  (F0/2) to satisfy the Nyquist frequency. However, based on the experiment presented here, this does not appear to happen as correlations (invariant to quotient frequencies) are very low in earlier layers.

\subsubsection{Formants}

To test how formants are encoded in the Generator network, we extract the first and second formant values  F1 and F2 (using script FormantPro by Xu and Gao \cite{formantpro} in Praat \cite{boersma15}). 

\begin{figure}
\centering
\includegraphics[width=.37\textwidth]{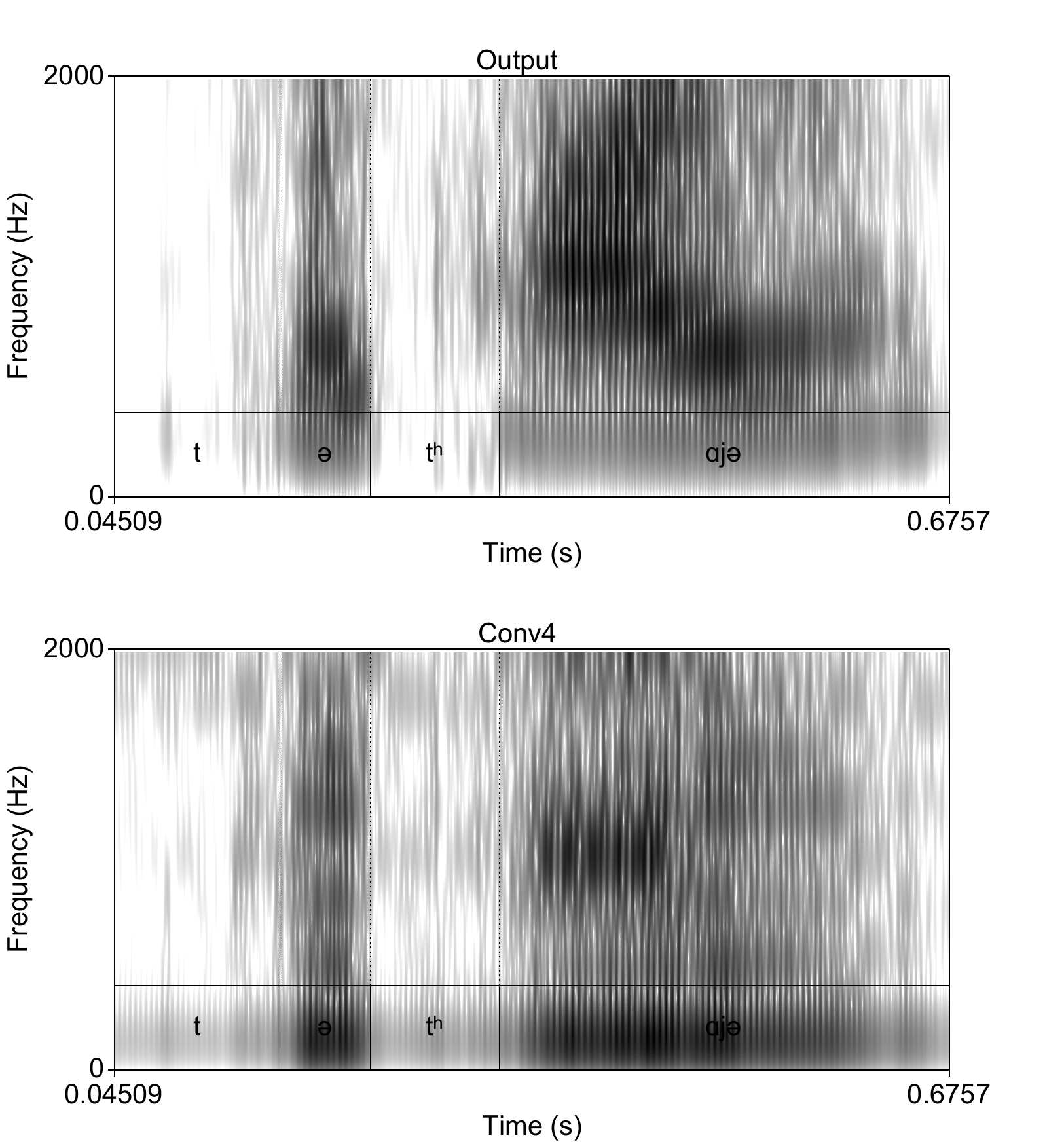}

\caption{Spectrograms (0-2000 Hz) of (top) the final generated output  of a reduplicated form [\textipa{ta"t\super hAj@}] (from the ciwGAN architecture when $c_1=0$ and $c_2=1$; transcribed by the authors) and (bottom) of the same reduplicated form with $c_1=0$ and $c_2=1$, but from the fourth convolutional layer (averaged across the feature maps after ReLU activation). }
\label{fig:spectrograms_tataja}
\end{figure}

The relationship in formant values between the output and  Conv4 is complex. First, formants are relatively challenging to estimate, even in clean human acoustic data, let alone in generated data or in intermediate convolutional layers. Second, while the fourth convolutional layer clearly features formant structure, the relationship between Conv4 and the final output is not straightforward. Figure \ref{fig:spectrograms_tataja} illustrates this relationship. The spectrogram of the output [\textipa{t@"t\super hAj@}] in Conv4 reveals a clear formant structure  (Figure \ref{fig:spectrograms_tataja}) but the actual formant values only partially overlap with the final output layer.

To quantify this observation, we analyze formant values  of the 38 periods with vocalic vibrations in normalized time and test the correlation between  the fourth convolutional layer and the final output.  The strongest correlation between the final output and the fourth layer appears to be in values of the second formant (F2): $r= 0.40$ 
(Figure 
22 in Supp.~Materials). In some outputs in the fourth convolutional layer (Conv4), F2 values match the final output layer both in the absolute values and in trajectories, but there also exist substantial deviations between the two layers. F2 is in a few cases already above the Nyquist frequency for Conv4 (2,048 Hz). F1, on the other hand,  does not appear to be faithfully encoded in Conv4: a correlation test between the output and Conv4 suggest a negative correlation for F1 ($r= -0.38$).

\subsection{Interpolation}
\label{2nd}

Results of the quantitative acoustic analysis of intermediate convolutional layers in Section \ref{sub1} through \ref{model2} reveal how and where the Generator encodes different acoustic properties. To interpret how linguistically meaningful representations in the latent space translate into spikes in activation in the intermediate layers, we use the proposal in Begu\v{s} \cite{begus19} and linearly interpolate individual latent variables to marginal levels well outside the training range.

\begin{figure*}
\centering
\includegraphics[width=.3\textwidth]{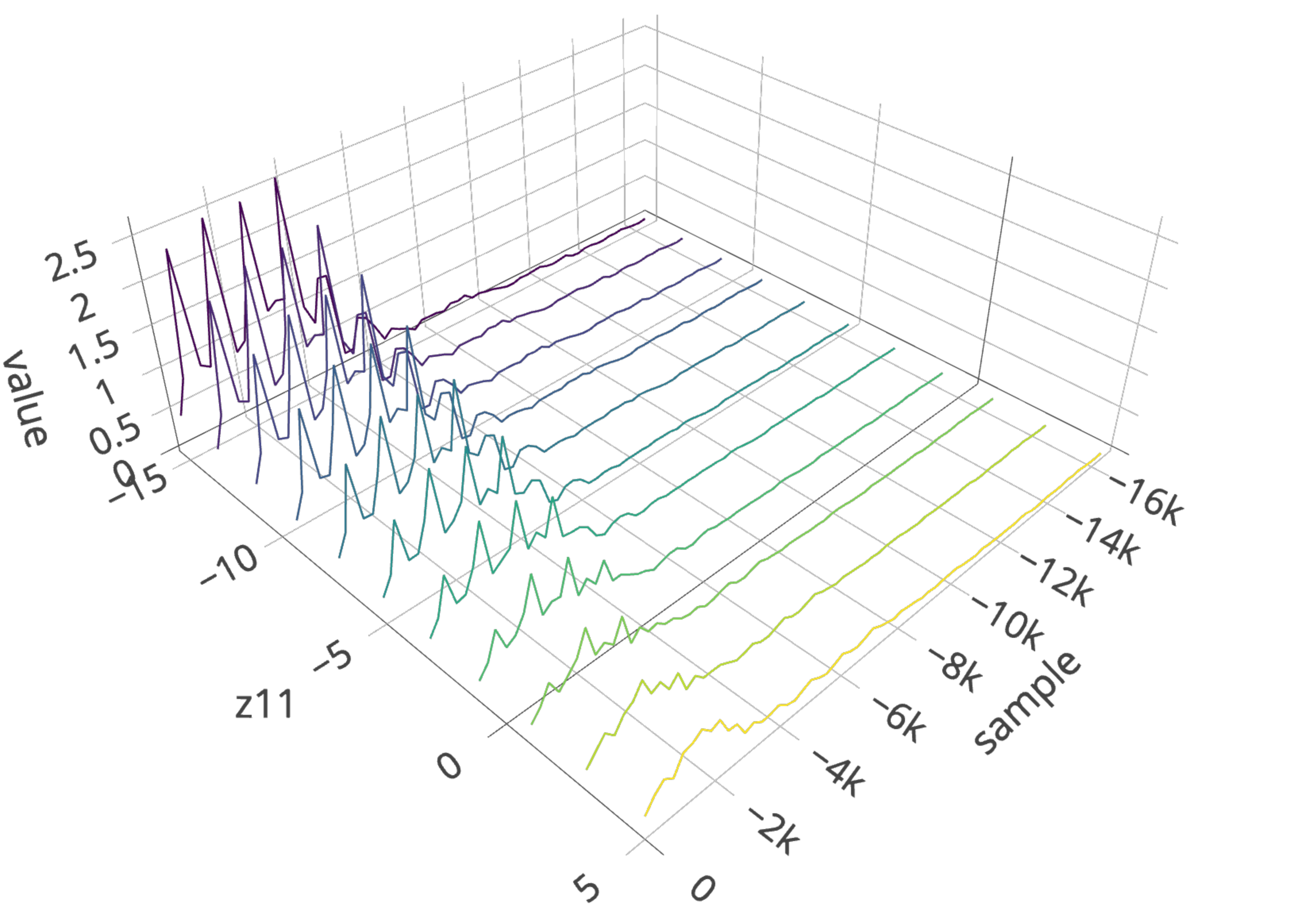}\includegraphics[width=.3\textwidth]{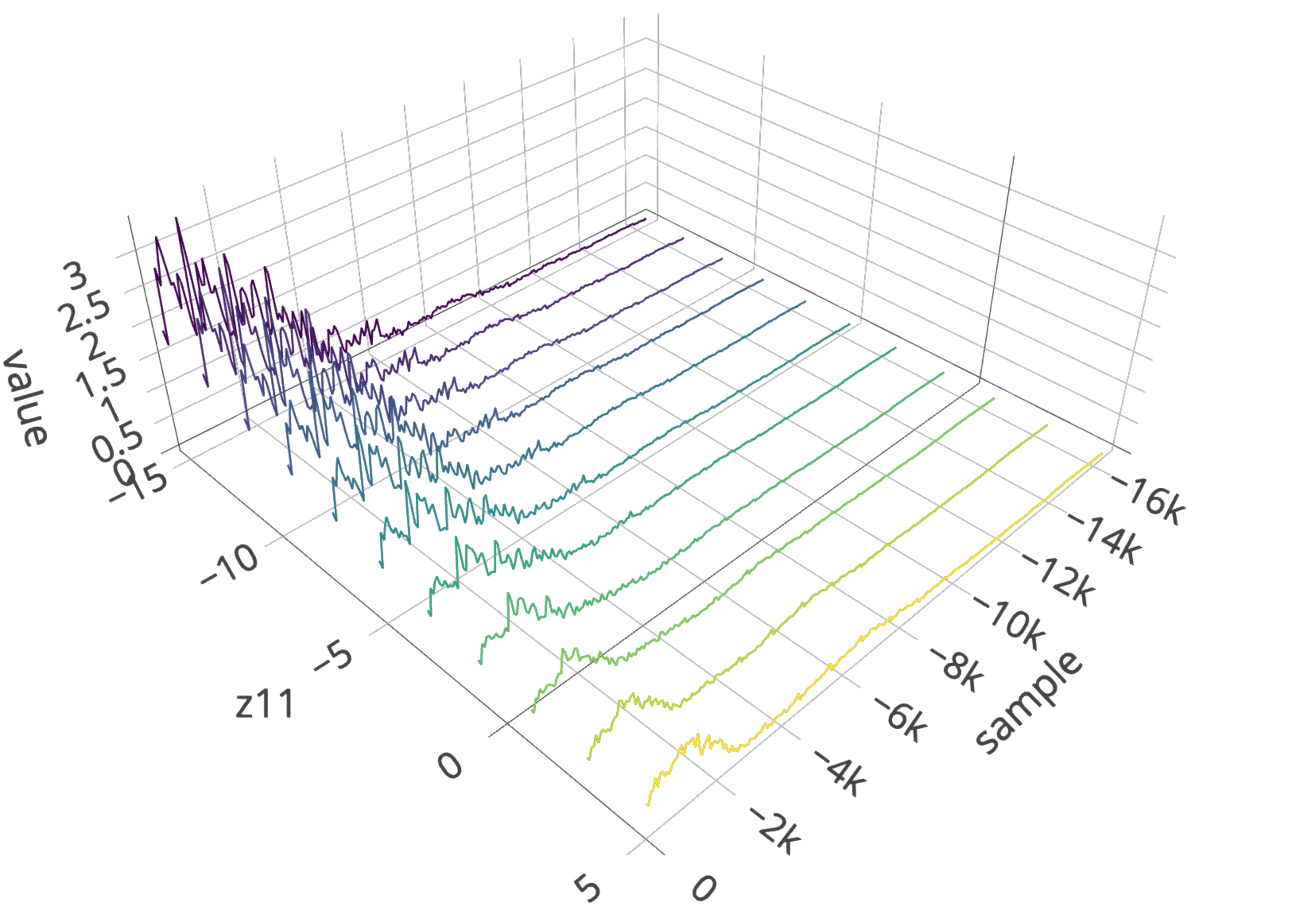}\includegraphics[width=.3\textwidth]{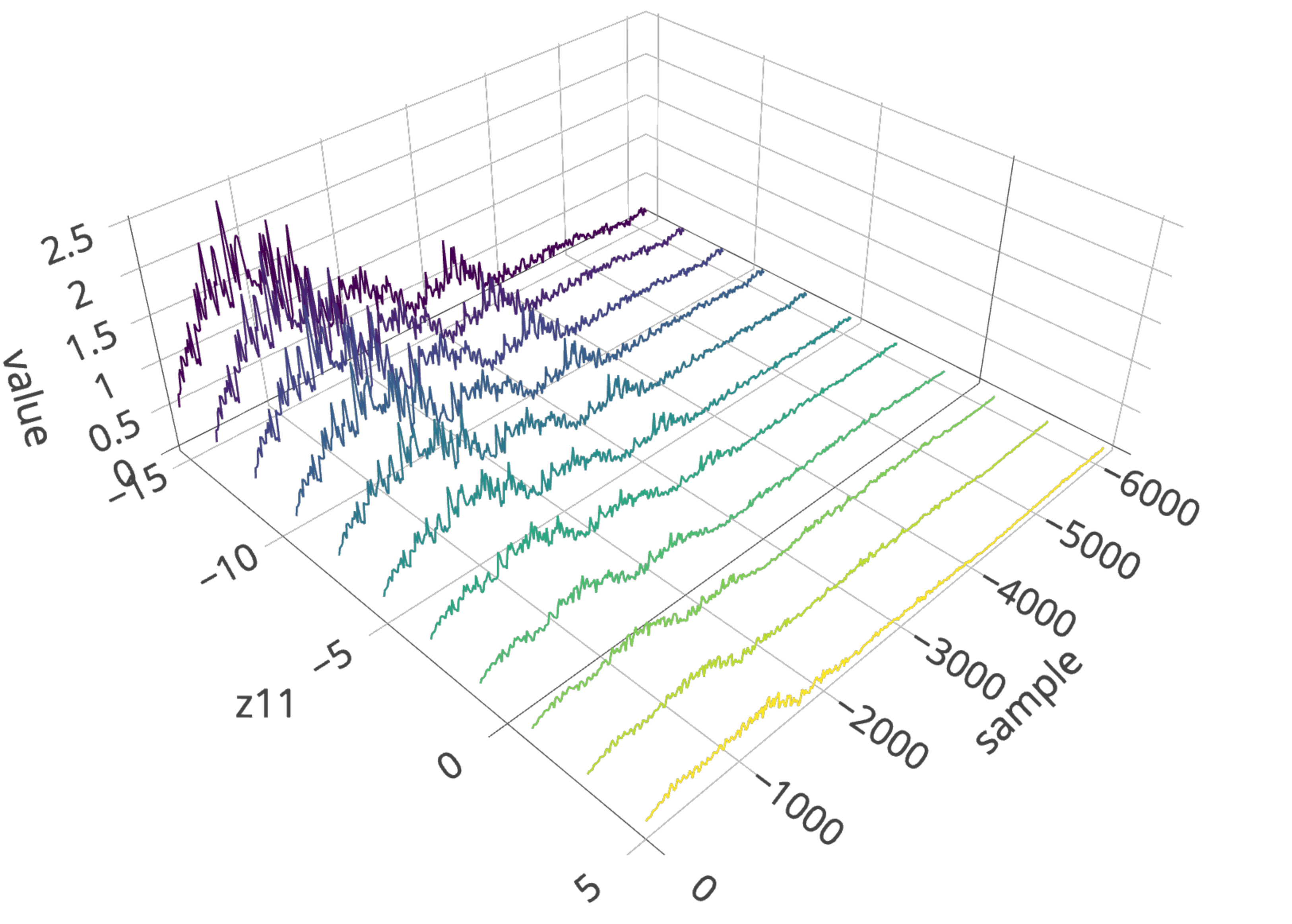}\\\includegraphics[width=.3\textwidth]{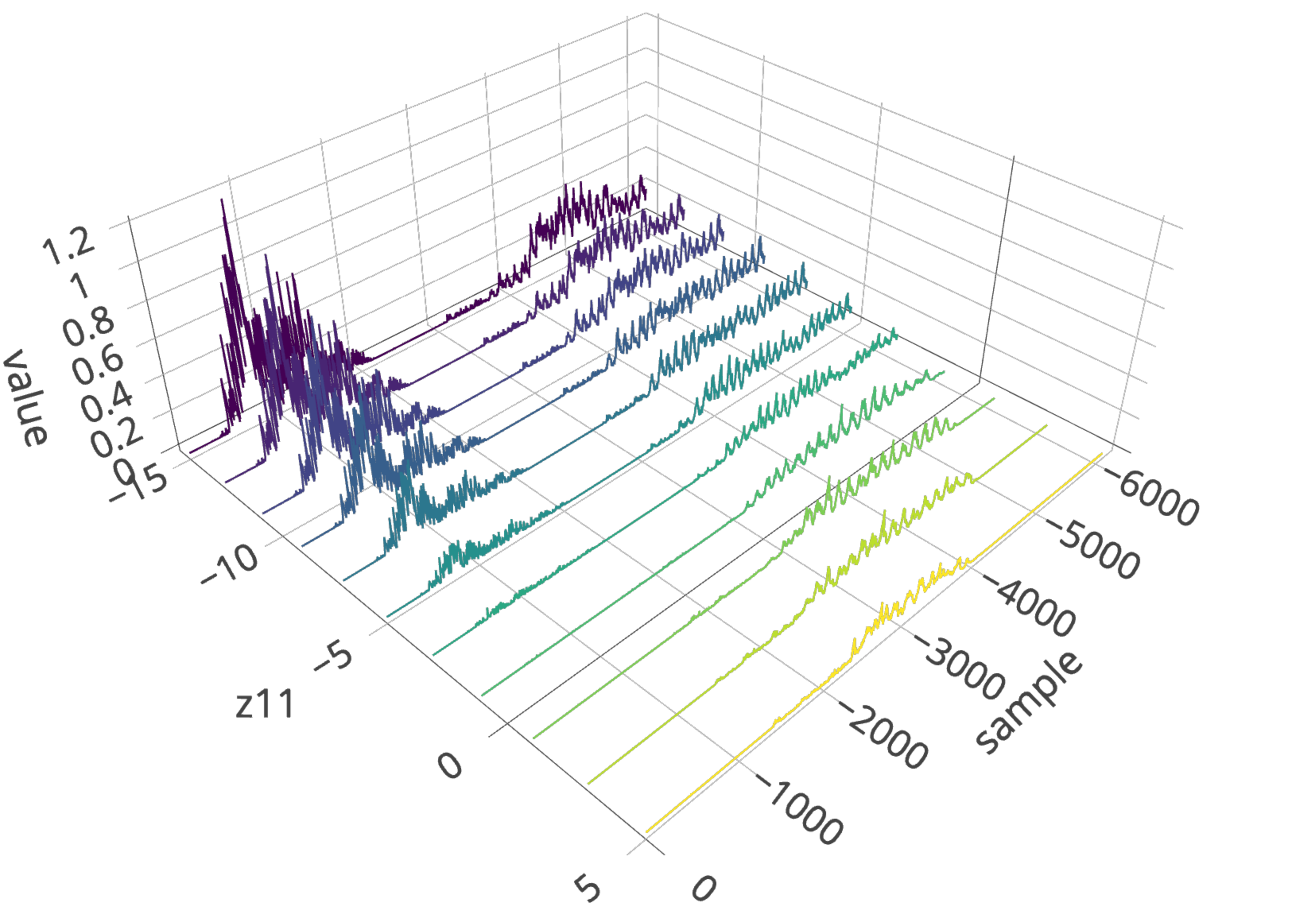}\includegraphics[width=.3\textwidth]{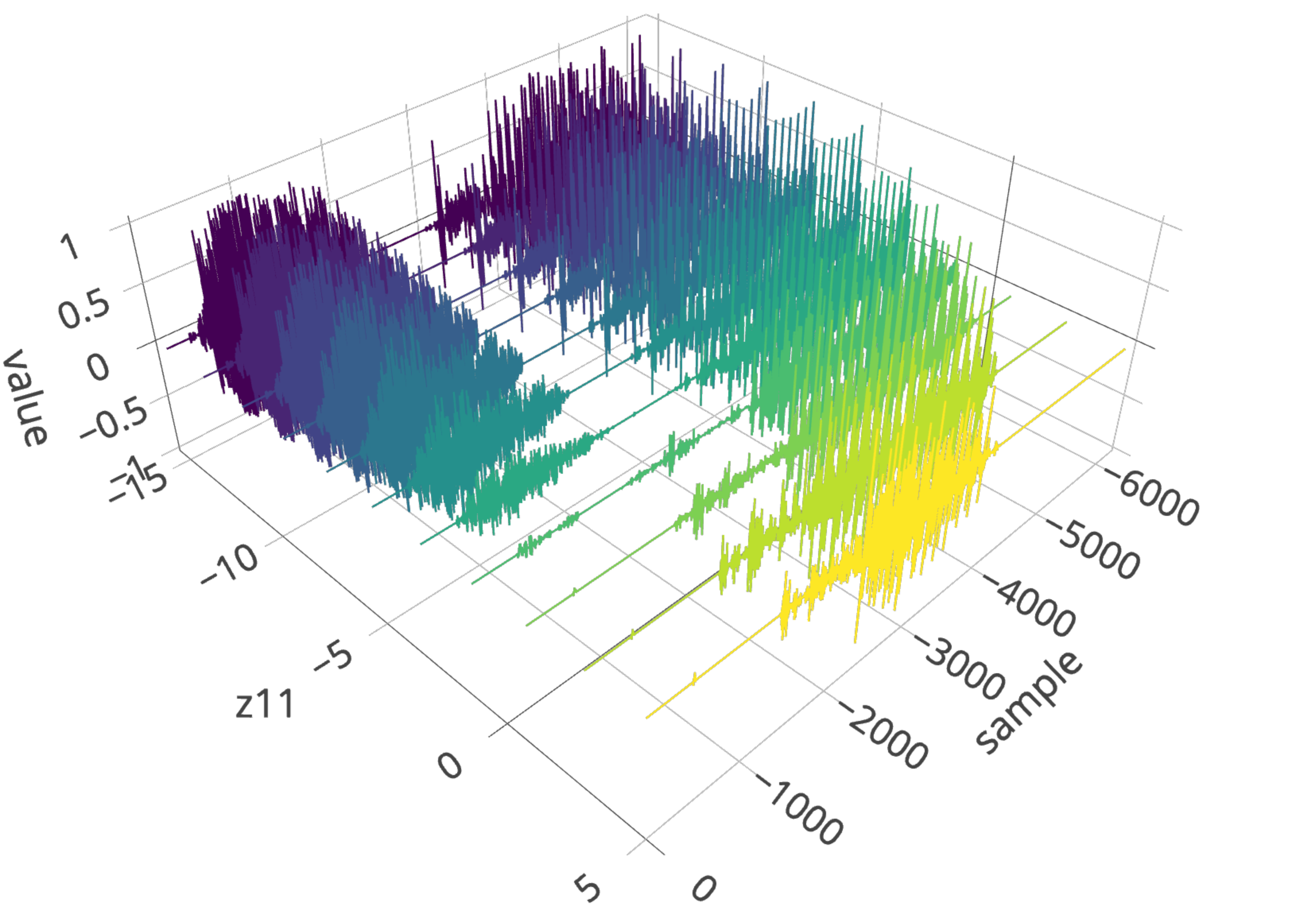}
\caption{Averaged values across feature maps after ReLU activation in the first (top left), second (top middle), third (top right), fourth (bottom left) convolutional layer, and the final waveform output (bottom right). For each convolutional layer, the graph represents 11 averaged values after ReLU activation where $z_{11}$ is linearly interpolated from -15 to 5 (with interval of 2) while all other 99 latent $z$ variables are held constant and limited to the training interval (-1,1) with uniform distribution. All outputs except in the final layer are upsampled with linear interpolation to total 16,384 samples (y-axis)  to match the audio waveform output. Representation of the third, fourth, and final layer were cut off at 6100th sample because higher samples featured mostly silence. The figures illustrate how linearly interpolating $z_{11}$ from 5 to -15 results in appearance of sound [s] in the final output and how representation of [s] is encoded across the layers.}
\label{fig:3dscatter}
\end{figure*}

We linearly interpolate values of $z_{11}$ in the bare WaveGAN model and values of the latent code $c_1$ and $c_2$ in the ciwGAN model. We generate outputs by linearly interpolating $z_{11}$ in the WaveGAN model from $-15$ to 5 (with interval of 2), and observe the resulting generated output for each value of $z_{11}$. This results in 11 outputs per each convolutional layer (55 total).   All other 99 latent variables remain constant across all outputs. The effects of this interpolation are similar across all sets. One such set of the five convolutional layers from the bare WaveGAN on TIMIT with interpolated values in the latent space is plotted in Figure \ref{fig:3dscatter}. The final output layer illustrates how an output without [s] gradually transforms into an output with [s] as $z_{11}$ is linearly interpolated towards the negative values which represent the presence of [s].

The advantage of the technique proposed in \cite{begus19} is that we can observe the causal effect of individual latent variables on the  output at each convolutional layer by analyzing averaged ReLU activations. Figure \ref{fig:3dscatter} illustrates how the linear interpolation of $z_{11}$ results in spikes of four values in the first convolutional layer. These four spikes increase as the  values of $z_{11}$ decrease, to the exclusion of other variables at this layer. It is likely the case that at the first layer (Conv1), the discretized abstract representation of [s] in the latent space transforms into spikes of a subset of values. At this point, the transformation is still highly abstract.  In the second convolutional layer (Conv2), the spikes transform into a more detailed representation of what corresponds to frication noise of [s] in the final output layer. The differentiation between the frication noise and periodic vocalic vibration becomes clearer in the third convolutional layer (Conv3). The increasing amplitude of the period corresponding to  frication noise (compared to the vocalic period) as the values of $z_{11}$ approach $-15$ suggests that the four spikes in values from Conv1 transform into precursors of frication noise and that linear interpolation of the individual latent variable $z_{11}$ representing [s] amplifies primarily the frication period throughout  the four layers and the final output. There is thus a causal relationship between $z_{11}$  and precursors of the  frication noise at each convolutional layer.  Visualization of the linear interpolations in the fourth layer (Conv4) also suggests that this layer encodes all major acoustic properties: frication noise, period of silence, and vocalic vibration as well as F0 and intensity of the periodic vocalic vibration.

\begin{figure*}
\centering
\includegraphics[width=.3\textwidth]{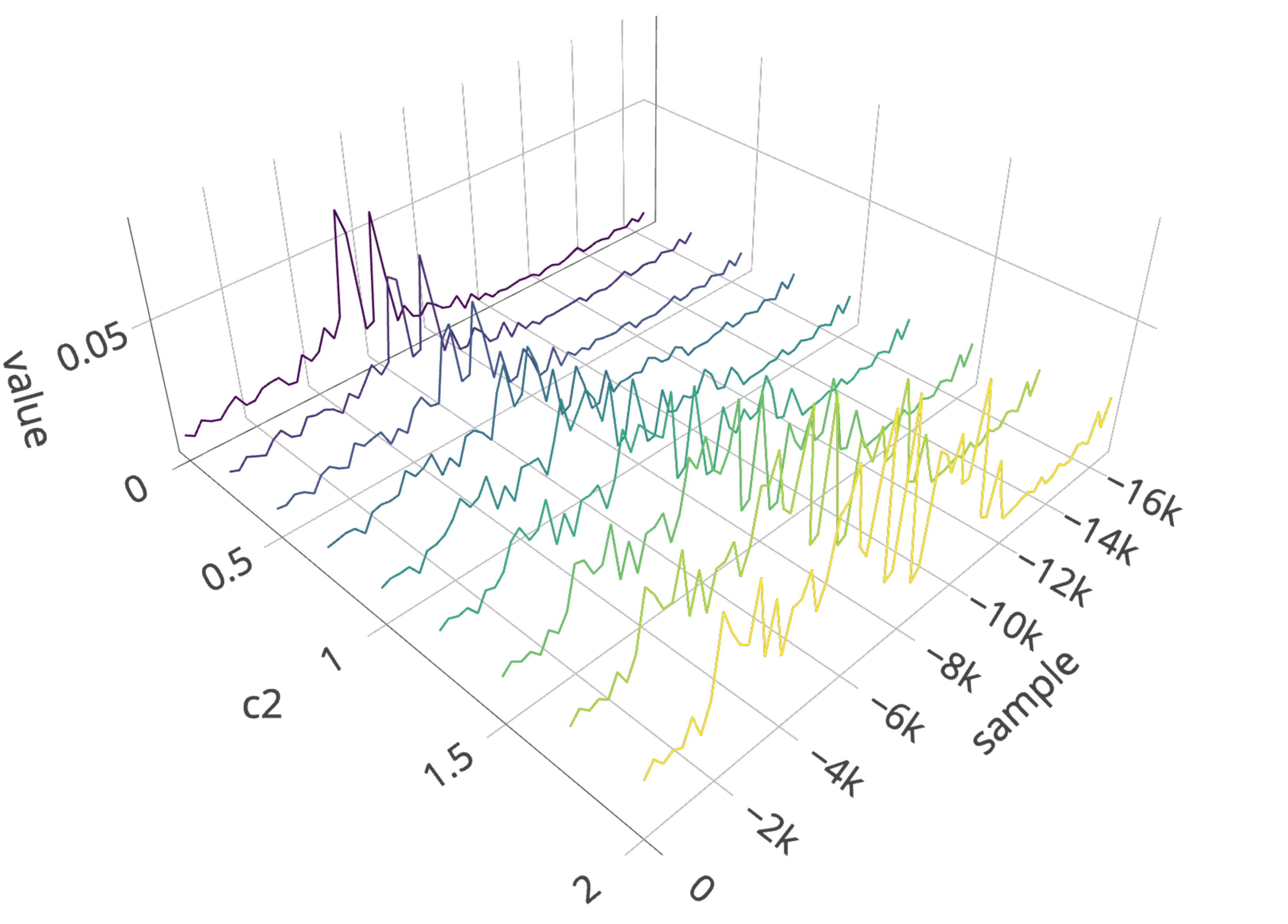}\includegraphics[width=.3\textwidth]{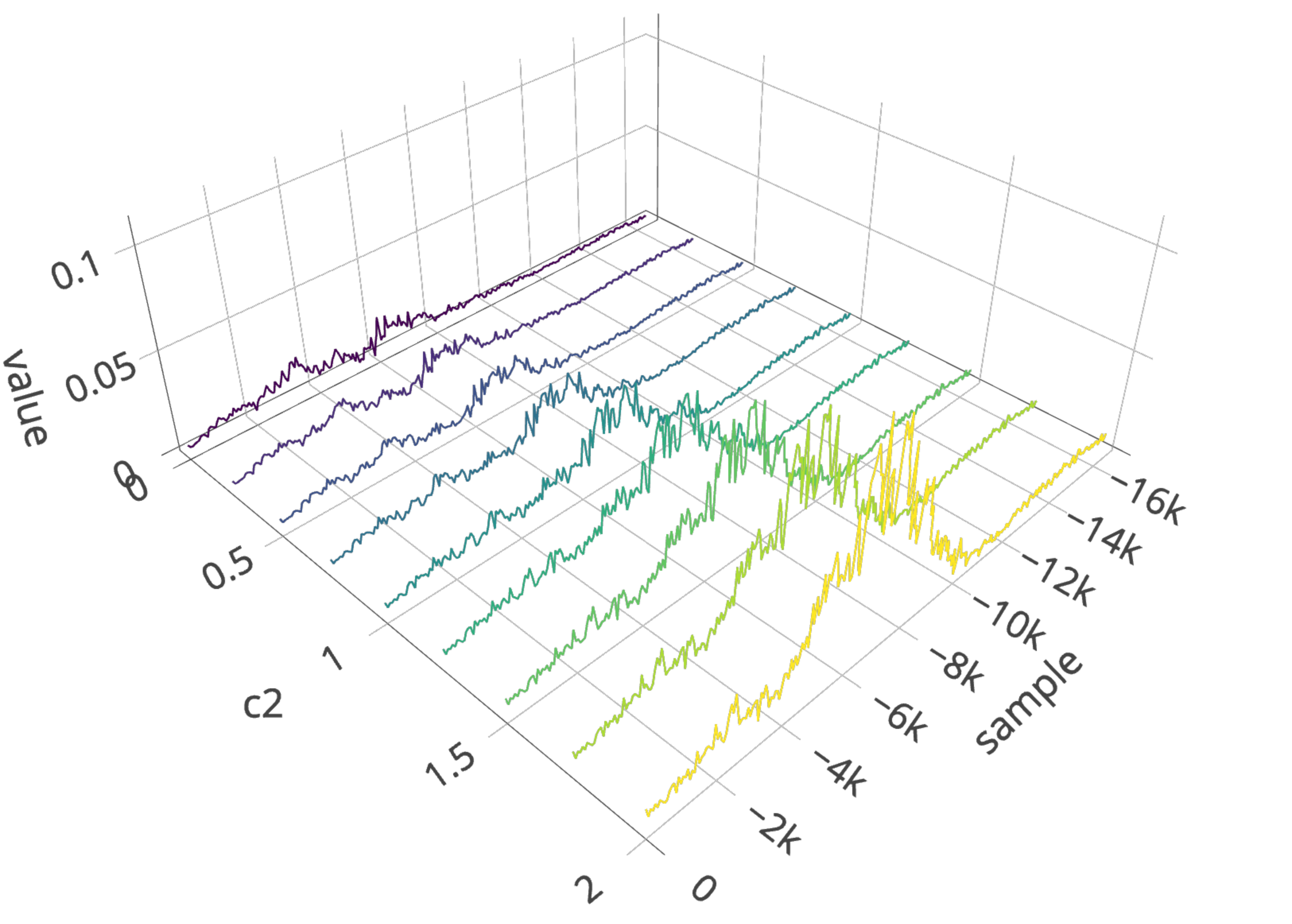}\includegraphics[width=.3\textwidth]{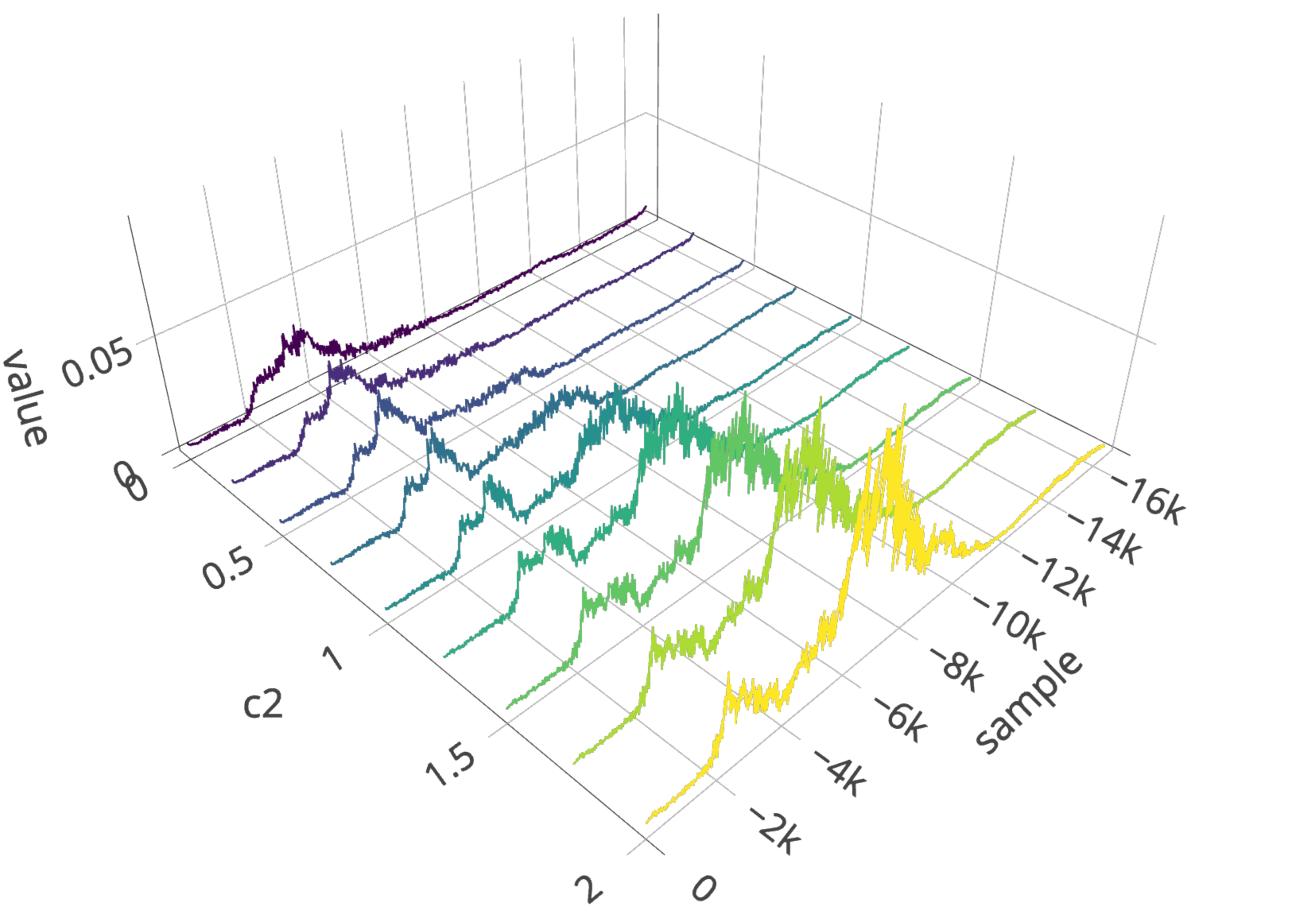}\\\includegraphics[width=.3\textwidth]{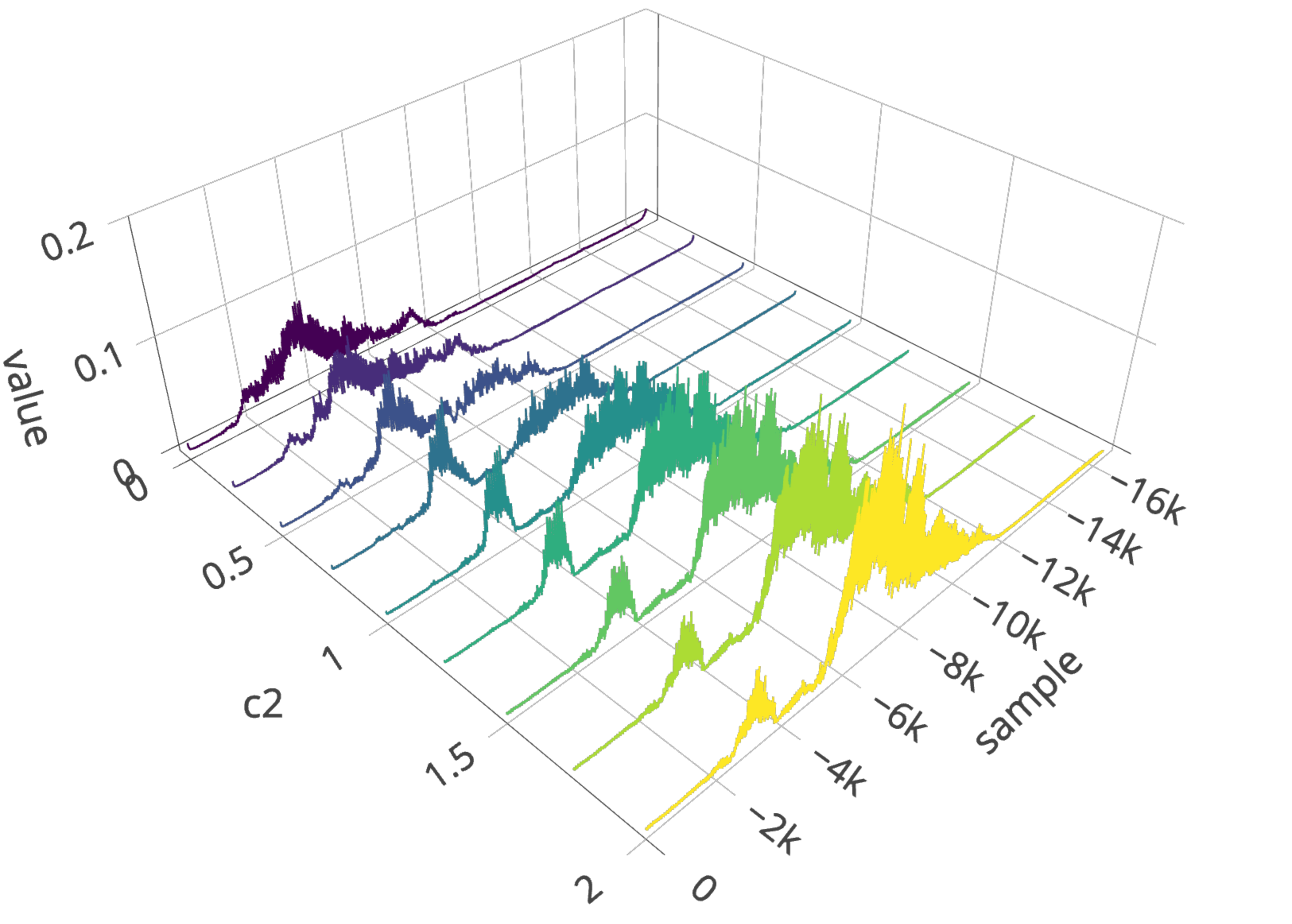}\includegraphics[width=.3\textwidth]{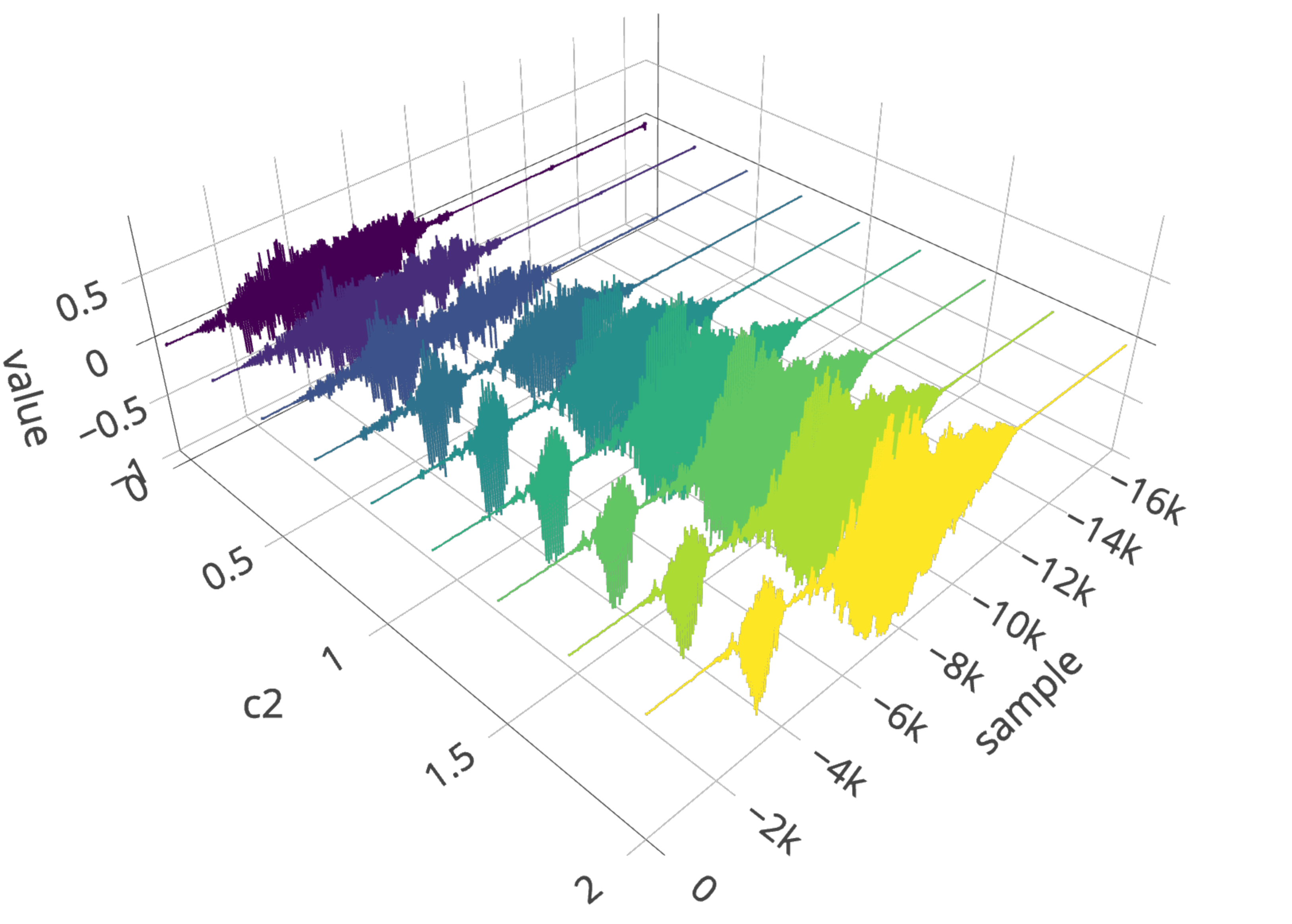}
\caption{Averaged values across feature maps after ReLU activation in the first (top left), second (top middle), third (top right), fourth (bottom left) convolutional layer, and the final waveform output (bottom right). At the values [0, 0] the final output layer can be transcribed as [\textipa{"dAji}]. At the values of the latent code [0.625, 0], the output can be transcribed as [\textipa{d@"daj}]; at the value  [1, 0] [\textipa{t@"t\super hAj@}]. For each convolutional layer, the graph represents 9 averaged values after ReLU activation where $c_{2}$ is linearly interpolated from 0 to 2 (with interval of 0.25) while $c_{1}$ is set to 0 and all other 98 latent $z$ variables are held constant and limited to the training interval (-1,1) with uniform distribution. All outputs except in the final layer are upsampled with linear interpolation to total 16,384 samples (y-axis)  to match the audio waveform output. Representation of the third, fourth, and final layer were cut off at 6100th sample because higher samples featured mostly silence. The figures illustrate how linearly interpolating $c_{2}$ from 0 to 2 results in appearance of reduplication and how reduplication is encoded across the layers.}
\label{fig:3dscatterCiw}
\end{figure*}

To interpret linear  latent code interpolation in the ciwGAN model trained on reduplication, we create a similar set: we manipulate the latent code from [0, 0] to [0, 2] with interval of [0, 0.25], thus creating 9 outputs per convolutional layer (45 total). One such set is chosen for visualization, but the effects of linear interpolation are similar across all sets.  All other 98 latent variables $z$ remain constant across all outputs.

Interpretation of linearly interpolated intermediate layers in the ciwGAN model is more complex because the phonological process the model is trained on --- reduplication (or copying) --- is computationally highly complex (see Section \ref{intro}). The first convolutional layer shows less discretized representations than in the \#sTV model. Interpolation from [0, 0] to [0, 2] (corresponding to presence of reduplication) seems to activate a few spikes for the main vowel and reduplicant vowel, but less categorically so than in the \#sTV model.  Averaged ReLU activations with linearly interpolated codes in Figure \ref{fig:3dscatterCiw} suggest that the latent code representing a computationally complex process results in the formation of two vocalic periods, interrupted by a consonantal element that appears identical on both sides of the reduplicated vowel (the copying principle). Visualizations also show that the period of silence (or reduced amplitude) corresponding to stop closure is encoded well into the third convolutional layer. Intensity (or acoustic envelope) appears to be encoded through all the convolutional layers.

\subsection{Individual feature maps and interpolation}
\label{individual}

\subsubsection{Individual feature maps}
\label{individual1}

In addition to the averaged feature maps at each layer, we also attempt to identify how linguistically meaningful properties are encoded separately in individual feature maps. Individual feature maps tend to be highly sporadic, with the same feature map possibly encoding different properties for distinct outputs even when the outputs have similar properties. However, there do exist some broad patterns across different generated outputs.

To identify these patterns for specific properties (such as presence of [s]), we generate a large number of outputs (N = 1000). Half of the outputs have been manipulated (using the latent space) so that the feature of interest ([s]) is present, and the other half have been manipulated so that the feature is absent. We perform this manipulation because the distribution of the Generator outputs without manipulation may have an uneven balance of the property we are interested in. All of the activations for these outputs are then averaged across each individual feature map. This creates a broad ``activation profile" for each filter across a variety of outputs. Clustering is then performed on these activation profiles to identify broad patterns of activation. 

\begin{figure*}
\centering
\includegraphics[width=0.9\textwidth]{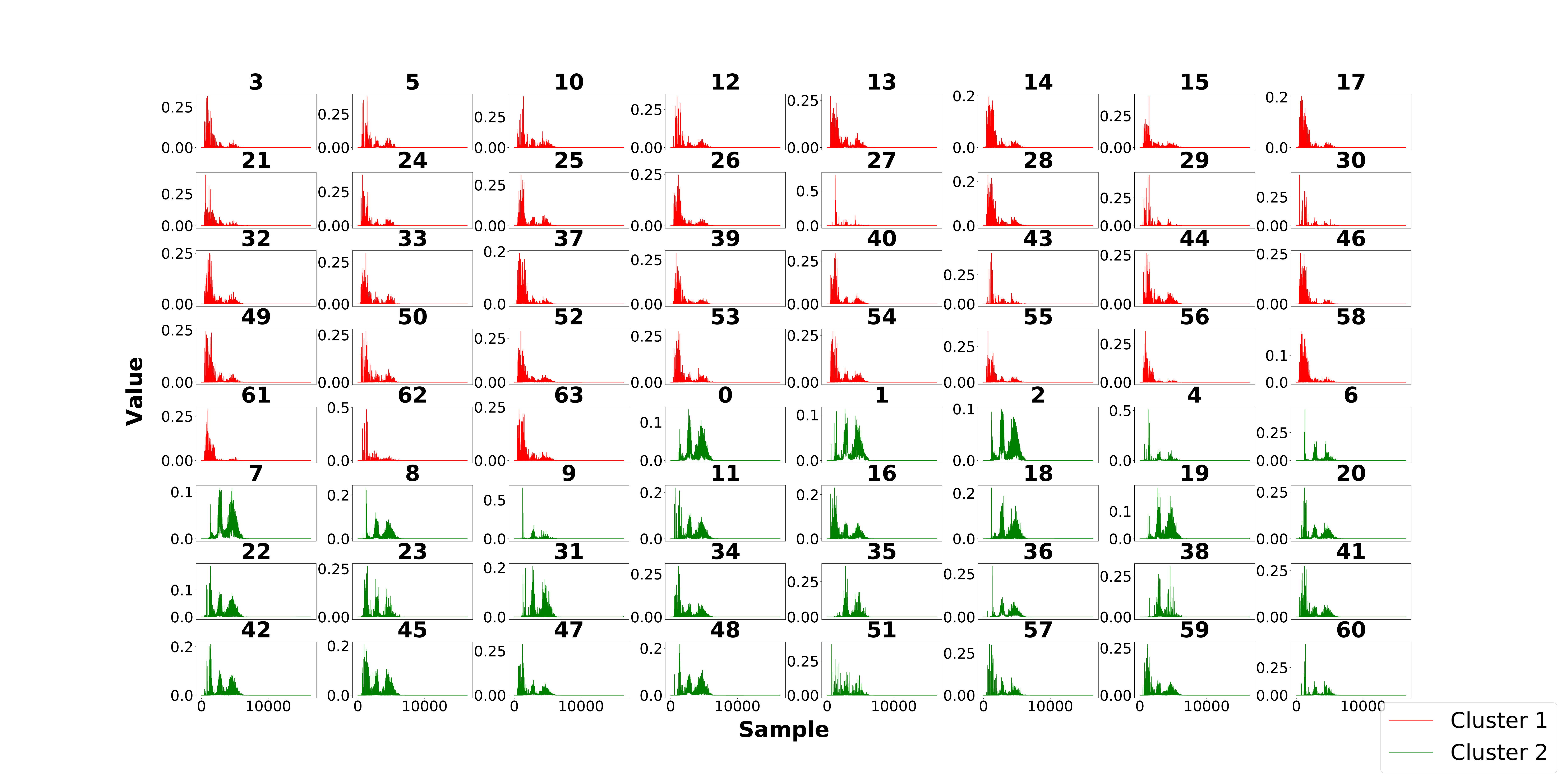}
\caption{Individual feature maps averaged over 500 instances of \#sTV and 500 instances of \#TV, clustered using spectral clustering. All feature maps exhibit an initial spike corresponding to the presence of a [s]-frication. However, the first cluster (red) has comparatively low activation after the initial spike, while the second cluster (green)  exhibits subsequent spikes that correspond to the rest of the sequence. These clusters were found with spectral clustering described in Section \ref{individual1}.}
\label{fig:clusters}
\end{figure*}

Specifically, we perform this analysis on the fourth convolutional layer of the WaveGAN model (Section \ref{bareganonan}), generating 1000 total outputs, 500 of which have $z_{11}$ set to -15, and 500 of which have $z_{11}$ set to 15. After averaging across feature maps, we perform spectral clustering, using the radius basis function kernel with a kernel coefficent of $1 \times 10^{-10}$ to construct the affinity matrix, and clustering using k-means where $k=2$. The results are shown in Figure \ref{fig:clusters}.

We see two broad distributions of activations: one in which there is a spike of activation near the beginning of the waveform and relatively low activation afterwards, and another in which we see additional spikes afterwards. We interpret the large single spike in the former category to correspond with the presence of a [s] frication, and determine these feature maps to encode almost exclusively for the presence of [s]. The latter category we take to also encode for [s], but which in addition is responsible for the rest of the \#sTV sequence. Indeed, when we average these clusters separately for particular examples of generated outputs with and without the [s] frication in Figure \ref{fig:clusters}, we see that the first cluster is activated weakly compared to the second in the absence of [s] ($z_{11} = 5$; Figure \ref{fig:clusters} top). In the presence of [s] ($z_{11} = -15$; Figure \ref{fig:clusters} bottom), we see activations from both clusters in the area corresponding to the [s]-frication, but weaker activations from the first cluster in the rest of the \#sTV sequence.

\begin{figure}
\centering
\includegraphics[width=0.4\textwidth]{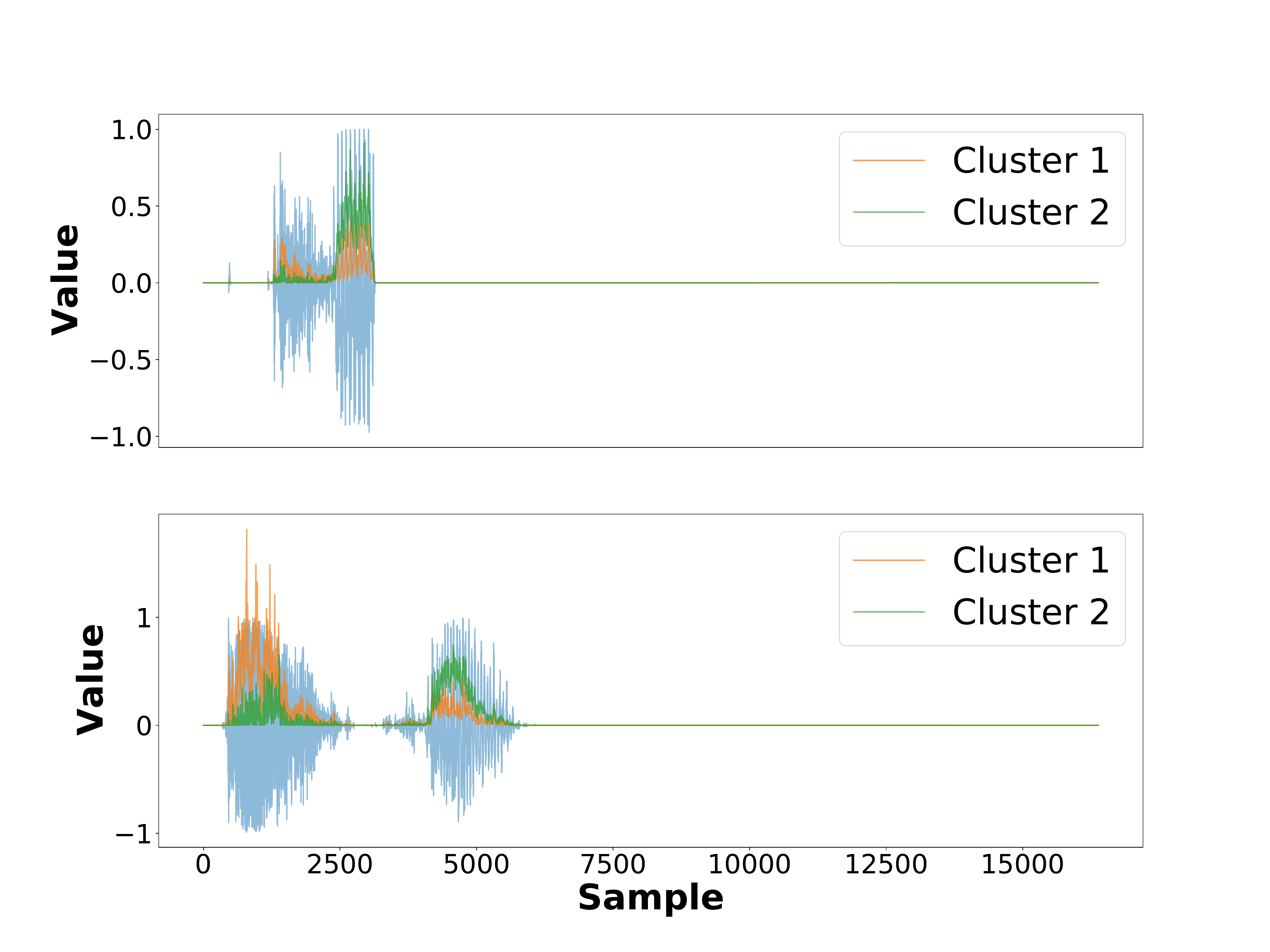}
\caption{The same clusters from Figure \ref{fig:clusters} averaged for a particular example of \#TV (top, $z_{11} = 5$) and \#sTV (bottom, $z_{11}=-15$), and plotted against the final output. Cluster 1 is much less activated than Cluster 2 in the \#TV output, but becomes highly activated in exactly the region corresponding to [s] in the \#sTV output.}
\label{fig:clusteraverage}
\end{figure}

\subsubsection{Interpolation} 

\begin{figure}
\centering
\includegraphics[width=.49\textwidth]{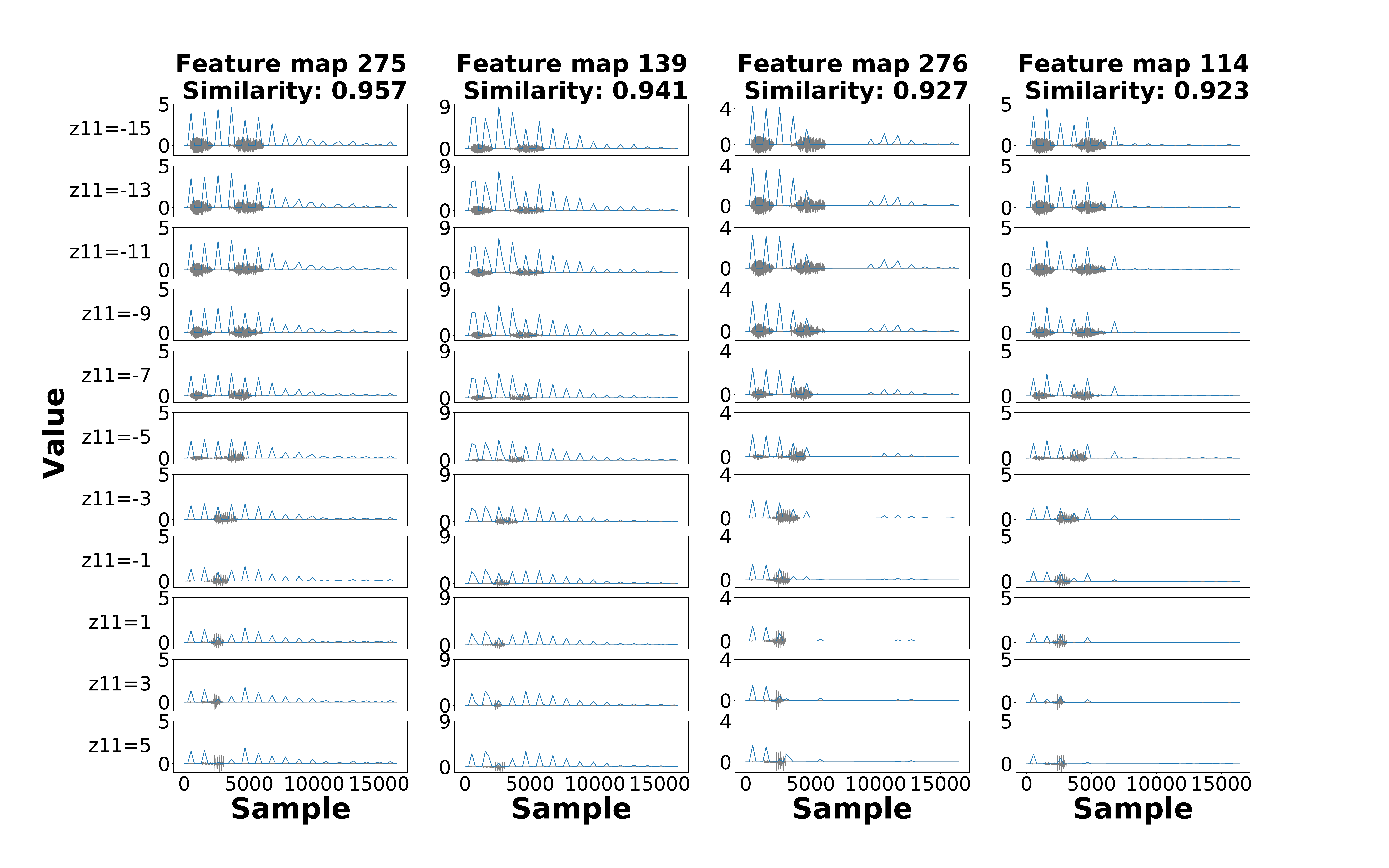}
\\
\includegraphics[width=.49\textwidth]{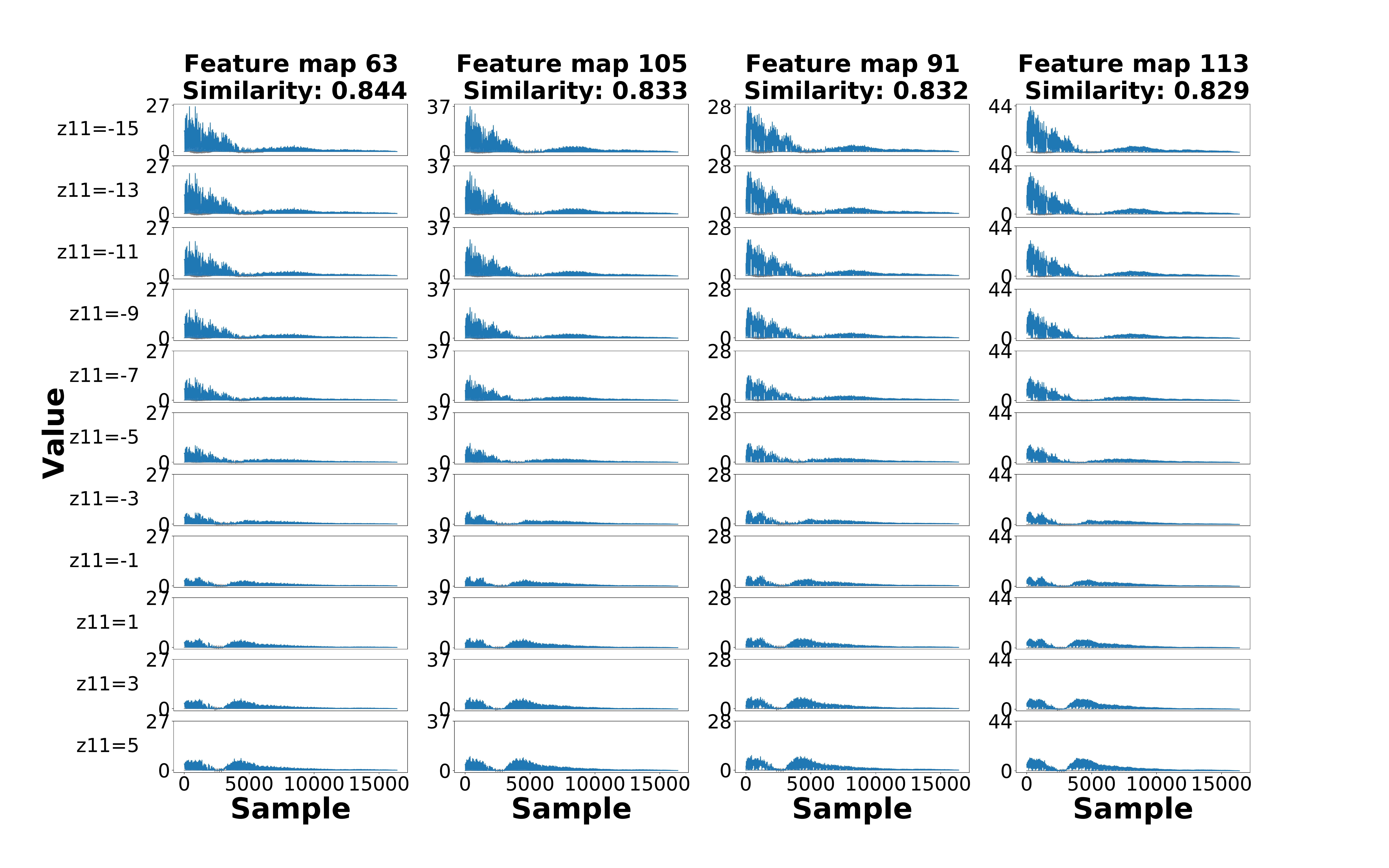}
\\
\includegraphics[width=.49\textwidth]{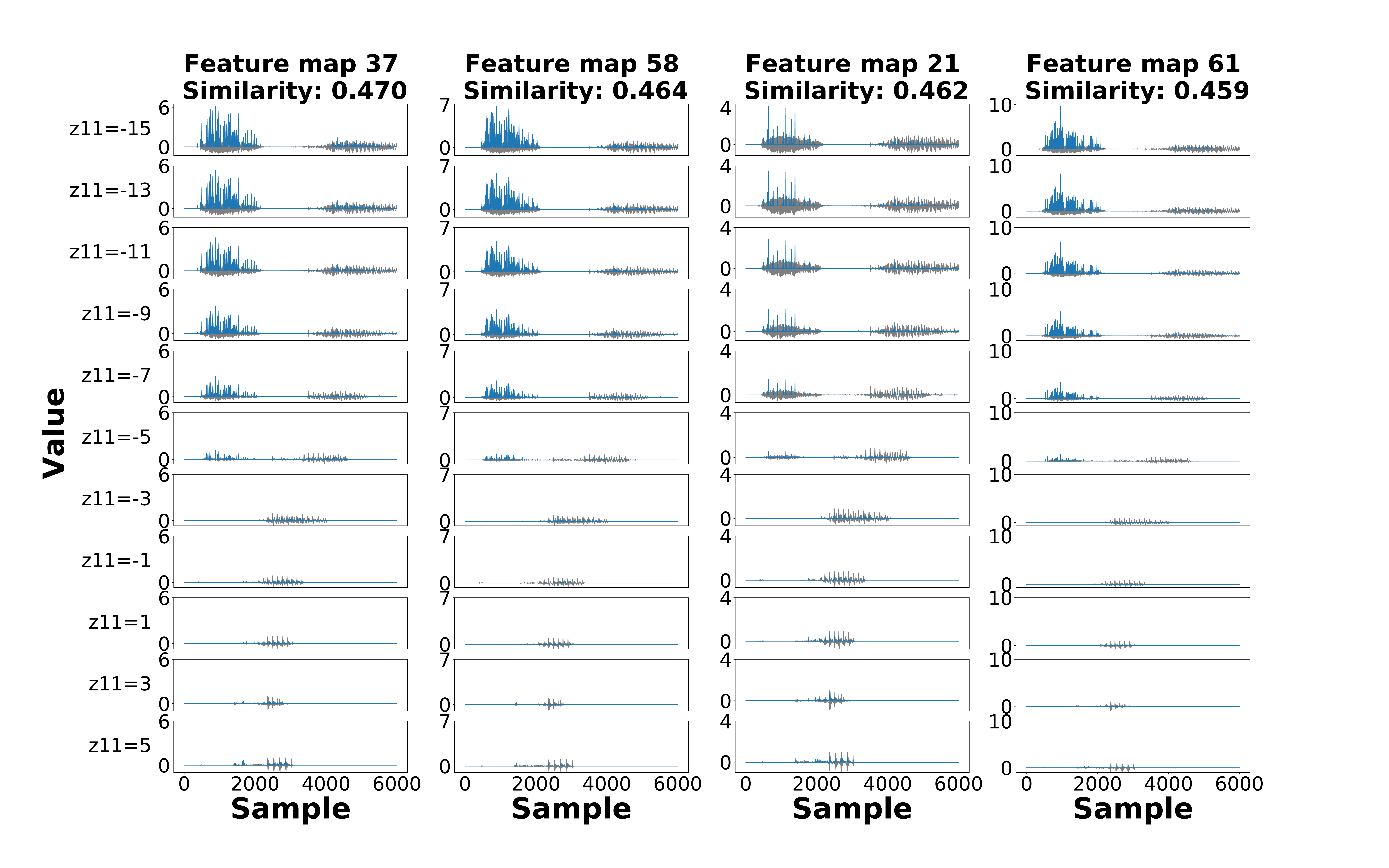}
\caption{Sets of individual feature maps after ReLU with minimal changes as determined by cosine distance from the values when $z_{11}=-15$. The feature maps are plotted at three convolutional layers: Conv1 (top), Conv3 
(middle), Conv4 (bottom).  Values of $z_{11}$ are linearly interpolated from $-15$ to 5 with interval of 2 for each convolutional layer and featue map (while other 99 $z$ variables are kept constant). }
\label{fig:individual}
\end{figure}

We can also analyze and interpret individual feature maps by linearly interpolating individual latent variables with linguistically meaningful representations. Figure \ref{fig:individual} illustrates four ``raw'' feature maps with linearly interpolated values of $z_{11}$ (in blue) and their corresponding final output layer (in gray). The four feature maps were chosen as those in which the distance between the feature map when $z_{11}$ is $-15$ and each corresponding feature map when $z_{11}$ is interpolated is smallest (according to cosine distance).

Individual feature maps show several parallels to the averaged values discussed in Section \ref{2nd}. By manipulating individual variables with linguistically meaningful representations (such as $z_{11}$), we can follow the causal effects of those variables on individual feature maps. Figure \ref{fig:individual} illustrates that individual feature maps transform marginal $z_{11}$ values into spikes in few values in Conv1. At Conv3, the $z_{11}$ transforms into a less abstract representation of frication noise  that substantially increases in amplitude as the values of $z_{11}$ approach $-15$. At Conv4, we see differentiation into periods of frication noise, silence, and periodic vocalic vibration. Again, linear interpolation results in increased amplitude of the frication noise.

Visualization of individual feature maps combined with linear 
interpolation of individual linguistically meaningful latent variables thus allows us to explore whether individual feature maps separately encode different phonetic properties (e.g.~frication noise, silence, or periodic vocalic vibration).

Exploration of individual feature maps (Figures \ref{fig:clusters} and \ref{fig:individual}) revealed no specific divergences in how linguistically meaningful units are encoded between individual maps and the summation of all values across the time domain, except that individual feature maps are highly sporadic if analyzed separately. This suggests that the summation technique proposed in this paper is representative of the properties that each layer encodes as a whole, and is useful for analyzing which acoustic properties are broadly encoded at which layers. Further work is needed to test whether individual maps encode additional information in higher layers and whether linguistically meaningful units are encoded not as time-series but as absolute values (for evidence of the latter, see \cite{begusZhou1}).

\section{Discussion}

This paper proposes a set of techniques to interpret and visualize outputs at intermediate transpose convolutional layers in CNN decoders (the Generator) trained on waveforms in an unsupervised manner. We argue that averaging across feature map values after ReLU activations  yields interpretable time series data that summarizes encodings of phonetic features at each convolutional layer in the Generator network. This allows us to use standard acoustic phonetic measurements to test what properties of speech are encoded at what layer.

\subsection{Acoustic properties across layers}

Acoustic analyses  suggest that many acoustic properties are encoded in the final convolutional layer before output (Conv4 or Conv9 in the deeper model). This layer features a clear period of frication noise (aperiodic vibration), a period of silence (corresponding to closure in voiceless stops) and a period of periodic vibration with some formant structure. Duration of the vocalic period is also  faithfully encoded in the final layer: periodic vibration between Conv4 and final output align almost perfectly. Visualizations in Figure \ref{fig:out}  suggest that timing of other major acoustic properties (frication noise and silence) is also highly aligned between Conv4 and final output.   Acoustic analysis of the final convolutional layer also suggests that  F0 and intensity values (or acoustic envelope) are faithfully encoded in this layer. 

Differences in the acoustic properties between the two models --- the bare WaveGAN and ciwGAN --- suggest that the degree to which individual acoustic properties are encoded at various intermediate layers can differ somewhat across the models. The two models probed here differ in the number of training steps (12,255 in WaveGAN vs.~15,920 in ciwGAN), the amount of training data  (5,463 total in the WaveGAN model vs.~996 total in the ciwGAN model), and consequently in the number of epochs (716 vs.~5,114). The structure of the Generator is identical across the models, except that in the ciwGAN architecture, the generator takes the latent code $c$ in addition to the latent variables $z$ as its input.  The ciwGAN model trained on a computationally more complex process with substantially more epochs appears to encode formant structure in the fourth convolutional layer (Conv4) more faithfully than the bare WaveGAN model trained on \#sTV.  While the relationship between the formant structure in Conv4 and the actual output is complex, the fourth convolutional layer does feature a clear formant structure which is at least partly correlated with the final output (in F2 values). 

The third convolutional layer in the 5-layer model is substantially more limited in what it can encode: with 1024 data points, its Nyquist frequency is 512 Hz. Formant structure is expectedly limited, but F0 and especially intensity data are encoded in this layer. 

Analysis of earlier layers (Conv1 and Conv2) and visualizations in Figure \ref{fig:3dscatterCiw} suggest that intensity (acoustic envelope) is attested well into the second and even first convolutional layer in the ciwGAN model. It appears  that the acoustic envelope gets encoded in the earliest layers and that acoustic information with higher frequencies (such as F0 and formant structure) is gradually built on top of the envelope in later layers.

\subsection{A causal relationship between the latent space and intermediate layers}

Combining the proposed interpretation technique with manipulation and linear interpolation of individual latent variables illustrates how individual variables in the latent space affect the activations at individual convolutional layers. Generating data with interpolated individual latent variables allows us to identify which activations in intermediate convolutional layers increase or decrease most substantially with interpolation, thus identifying a causal relationship between individual latent variables and activations in intermediate layers.

We can also probe individual feature maps by manipulating and linearly interpolating individual latent variables. The effects of linear interpolation on individual feature maps is similar to its effect on the averaged values (Section \ref{individual}). 

Analysis of individual feature maps by manipulation of the latent space also suggests that different acoustic features (such as aperiodic frication noise or periodic vocalic vibration) can be encoded in separate feature maps. Clustering in Section \ref{individual1} suggests that some feature maps activate the frication part more strongly when the latent variable corresponding to [s] is manipulated to marginal levels, while in others the vocalic period is activated more strongly. 

\subsection{Applications, limitations, and future directions}

With the proposed technique, we can analyze which acoustic properties are encoded  in which intermediate convolutional layers. ASR and speech synthesis systems overwhelmingly include convolutional neural networks, at least in initial layers. Recently, there has been a shift towards using ASR/synthesis systems directly from raw waveforms \cite{zeghidour18}. Additionally, ASR increasingly involves unsupervised models, and recently a GAN-based approach has been proposed for unsupervised ASR with no labeled data required \cite{baevski21}. Our proposal allows visualization and interpretation of transpose convolutional layers in a GAN-based unsupervised model that operates from raw waveforms. While most CNNs in ASR systems involve windows shorter than 1 s (as in our case), we choose to apply the proposed technique to longer windows in order to test the encoding of not only acoustic properties, but also of higher-level phonological processes (such as reduplication). Understanding how phonological processes are encoded will be increasingly important as unsupervised speech technology systems become available in languages with substantially more (and more complex) phonological processes than English. Finally, our proposal allows exploration of the causal relationship between individual latent variables and intermediate convolutional layers by manipulating and linearly interpolating latent variables to values outside of the training range. Exploring causal relationships in deep learning models is a growing area in machine learning research.

We apply the proposed technique to two GAN models trained on limited and curated data, because the latent space can be highly interpretable in GANs \cite{begus19,begusCiw,begusIdentity}. We also limit our discussion on the Generator (decoder) network (for interpretation of the classifier network, see \cite{begusZhou1}). These GAN-based models, while capable of both speech synthesis and speech classification \cite{begusCiw}, are not usually employed in most current ASR/synthesis applications. The results from the deeper model (Section \ref{deepbareganonan}) trained on LibriSpeech, however, suggest that the proposed technique can be scaled to larger models and that similar (but more distributed) encoding emerges in intermediate layers of deeper models as well. Future directions should involve applying the proposed technique to ASR and TTS models that involve CNN layers (e.g.~wav2vec 2.0 \cite{baevski20}). Similar approaches can also be utilized on any CNN-based decoder (such as \cite{ping18}) as well as on VAEs, which are often used in unsupervised speech technologies \cite{chung16,kamper19,chorowski19,niekerk20}. Like GANs, VAEs involve upsampling from a latent space (whether distributed or as a codebook) with a decoder model similar to a Generator. The combination of intermediate convolutional layer visualization and interpolation of individual variables can provide insights into learning in VAEs as well.

In this paper, we also limit our discussion on the most salient acoustic properties (intensity, F0, and formant structure). Other properties such as acoustic correlates of gender, dialects, race, or socioeconomic background can be probed with the same techniques as well.

The interpretation and visualization technique can serve also as a diagnostic for improving the performance of CNNs trained on speech.  The interpretation suggest that several acoustic properties relevant to speech perception (especially the formant structure of vowels) is encoded only in the final layer (of the 5-layer models), primarily because the Nyquist frequency does not allow properties with higher frequencies to be directly encoded as a time-series property (i.e.~with frequency encoding) earlier in the structure of the Generator network. This suggests that introducing more layers capable of encoding properties with higher frequencies might improve performance of the model. Testing this hypothesis is left for future work. The proposed technique can also be applied to unsupervised acoustic classifiers. \cite{begusZhou1} apply it to intermediate layers in the Q-network and additionally propose that both shapes and absolute values of learned representations can be inferred with non-linear regression.

The proposed technique can also serve for direct (albeit superficial) comparisons between intermediate convolutional layers and neural activity in the brain \cite{begusZhouZhao}. A few parallels are immediately available: the output at the fourth convolutional layer (Conv4) resembles the complex auditory brain stem response when subjects are presented with acoustic vocalic stimuli (as in \cite{zhao18}). Also, parallel to the intensity values (or acoustic envolope) which are encoded high in the structure of the convolutional network (up to the second and even first convolutional layer in the ciwGAN), the acoustic envelope is encoded relatively high in the brain as well (in the auditory cortex; \cite{oganian19}). The advantage of the proposed technique is that it outputs time-series data and enables testing of which acoustic properties are encoded at which layers. This information  can be used for comparison between the convolutional networks and various neuroimaging techniques (which also output time-series data).

\ifCLASSOPTIONcaptionsoff
  \newpage
\fi



\bibliographystyle{IEEEtran}
%

\bibliography{bib.bib}



%







\end{document}